\documentclass[pra,twocolumn,letterpaper,superscriptaddress]{revtex4}
\usepackage{graphicx,amsmath,amssymb,amsfonts,latexsym,color,dcolumn,bm,epsfig,subfigure}
\usepackage[plainpages=false,hyperfootnotes=false,colorlinks=false]{hyperref}
\usepackage[normalem]{ulem}
\usepackage{dsfont}
\usepackage{array}
\usepackage{pgfplots}
\usepackage{here}

\renewcommand{\imath}[0]{\mathrm{i}}

\begin{document}

\title{Nonequilibrium atom-surface interaction with lossy multi-layer structures}%

\author{Marty Oelschl\"ager}%
\affiliation{Max-Born-Institut, 12489 Berlin, Germany}

\author{Kurt Busch}
\affiliation{Max-Born-Institut, 12489 Berlin, Germany}
\affiliation{Humboldt-Universit\"at zu Berlin, Institut f\"ur Physik, AG Theoretische Optik \& Photonik, 12489 Berlin, Germany}

\author{Francesco Intravaia}
\affiliation{Max-Born-Institut, 12489 Berlin, Germany}

\affiliation{Humboldt-Universit\"at zu Berlin, Institut f\"ur Physik, AG Theoretische Optik \& Photonik, 12489 Berlin, Germany}

\begin{abstract}
The impact of lossy multi-layer structures on nonequilibrium atom-surface interactions is discussed.
Specifically, the focus lies on a fully non-Markovian and nonequilibrium description of quantum
friction, the fluctuation-induced drag force acting on an atom moving at constant velocity and
height above the multi-layer structures. Compared to unstructured bulk material, the drag force
for multi-layer systems is considerably enhanced and exhibits different regimes in its velocity
and distance dependences. These features are linked to the appearance of coupled interface polaritons
within the superlattice structures. Our results are not only useful for an experimental investigation
of quantum friction but also highlight a way to tailor the interaction by simply modifying the
structural composition of the multi-layer systems. 
\end{abstract}
\maketitle
%\tableofcontents
\section{Introduction}
The notion of \textit{vacuum} changed dramatically after the rise of quantum mechanics. The existence
of quantum fluctuations even at zero temperature and the possibility of ``structuring the vacuum''
has led to the discovery of many new interesting phenomena.
In this context many vacuum fluctuation-induced interactions such as the Casimir and Casimir-Polder
effect \cite{Casimir1948} have been investigated. Strongly related to the Casimir-Polder force, is quantum
friction, a drag force that even at zero temperature opposes the relative motion of two or more objects
\textit{in vacuum} \cite{Intravaia2016}.
One of the most studied configurations consists of an atom (or a microscopic object) moving parallel
to a surface at constant height and velocity \cite{Intravaia15,Pieplow15,Dedkov02a,Scheel2009}.
In such a system, quantum friction has a simple interpretation
in terms of the interaction between the moving microscopic object and its image within the
material below. The motion of the image is ``delayed" due to the frequency dispersion of the material
permittivity, leading to both a modification of the equilibrium Casimir-Polder force perpendicular to
the surface and a component of the force parallel to the surface that opposes the motion.
Already in this simple picture, we can intuitively understand the relevance of two mechanisms at work
in the quantum frictional process: The strength of the coupling between the microscopic object and its
image and the resistance felt by the image when dragged through the material.
The latter can be related to the dynamics of the charge carriers within the material composing the
substrate and, in particular to its resistivity.
Instead, the coupling strength strongly depends on the 
electromagnetic densities of states characterizing the system. Altering either one of these aspects 
will eventually lead to an modification of quantum friction.
A currently popular class of systems where this can be implemented are nanostructured substrate materials 
\cite{Yan2012,Liu2007,Shekhar2014}. 
In the framework of fluctuation-induced forces, they have already been considered both in theoretical 
analyses (see for example \cite{Messina2017,Lambrecht2008,Rodriguez11,Intravaia12a,Intravaia15}) and in experiments 
\cite{Chan08,Intravaia13,Bender14,Tang17,Chan18}.
One prominent example is given by nanoscaled multi-layer structures, where a specific pattern of 
distinct layers are repeatedly stacked forming a superlattice
\cite{Saarinen2008,Poddubny13,Guo14b}. 
When carefully designed, they are known for exhibiting effective hyperbolic dispersion relations 
\cite{Iorsh12,Kidwai12,Poddubny13}, 
which have applications in many fields of research \cite{Belov06,Wurtz11,Cortes12}.
\begin{figure}
\centering
\includegraphics[width=0.4\textwidth]{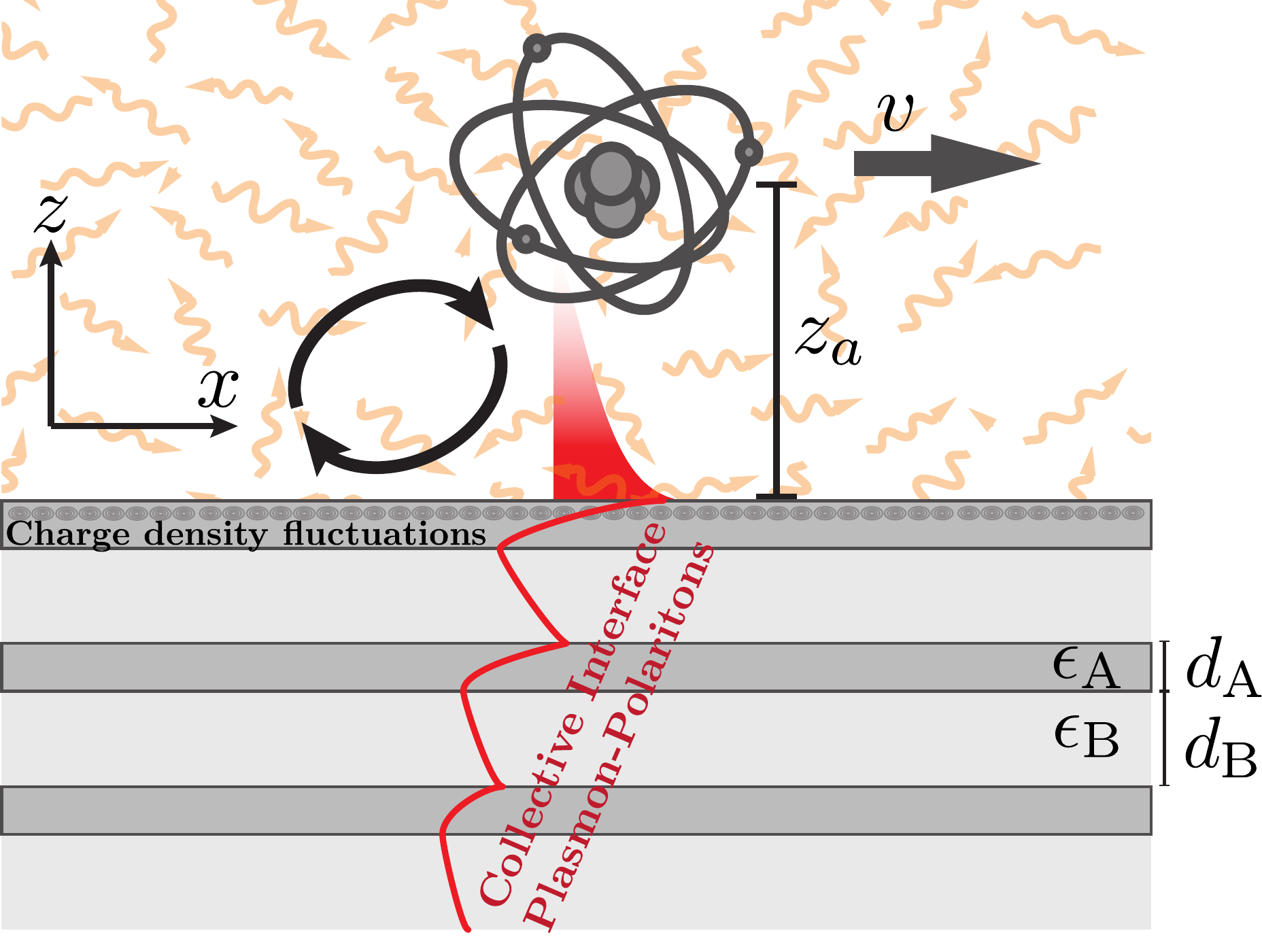}
\caption{A schematic description of the system considered in this work. An atom (or microscopic
         object) moves at constant velocity and constant height above a half-space made by 
				 periodic sequence of alternating conductive ($\epsilon_{\rm A}$) and dielectric 
				 ($\epsilon_{\rm B}$) layers with corresponding thicknesses $d_{\rm A}$ and $d_{\rm B}$. 
				 The spectrum of vacuum fluctuations is "stuctured" through the properties of the 
				 multi-layered structure and gives rise to a nonequilibrium atom-surface interaction 
				 which opposes the motion of the atom. 
				 This quantum frictional force is affected by the appearance of electromagnetic 
				 resonances due to the inter-layer interaction of plasmon-polaritons at the 
				 dielectric-metal interfaces and can be tailored by acting on the geometry and 
				 the material properties of the individual layers.
\label{fig:setup}}
\end{figure}
In the present work, we investigate how the characteristic behavior of quantum friction 
and specifically its strength and functional dependence on the velocity and position of 
the atom, is modified when the planar medium is a two-component superlattice of alternating 
layers with metallic and dielectric properties. The paper is organized as follows.
In Section \ref{Sec:theoryQF}, we briefly review the theory of quantum friction, highlighting 
the features connected with the properties of the surface, which can lead to a modification 
of the interaction through nanostructuring. We introduce in Section \ref{Sec:supEMA} our 
material models, analyzing the properties and the physical parameters which are relevant 
to the quantum frictional process.
Finally, in Section \ref{Sec:QFres}, we merge the insights of the previous two sections 
and explicitly calculate the quantum frictional force on an atom moving above
a superlattice.

\section{Quantum Friction}
\label{Sec:theoryQF}

Physically, quantum friction can be derived from the Lorentz force: If we choose the $z$-axis to 
be perpendicular to the surface (see Fig.~\ref{fig:setup}), the force lies in the $(x,y)$-plane 
against the direction of motion. This means that if the atom moves at constant height $z_a$ above a flat surface with constant velocity $\mathbf{v}$, then $\mathbf{F}=F\mathbf{v}/v$ ($v=|\mathbf{v}|$) \cite{Intravaia2016}. In our description,
the atom is described in terms of a time-dependent dipole operator $\hat{\mathbf{d}}(t)$: For 
simplicity, we further assume a rigid dipole configuration $\hat{\mathbf{d}}(t)=\mathbf{d}\hat{q}(t)$, 
where $\mathbf{d}$ is the static dipole vector and $\hat{q}(t)$ describes the dipole's
internal dynamics. 
For systems at temperature $T=0$, it has been shown~\cite{Intravaia2014,Intravaia2016} 
that the quantum frictional force $\mathbf{F}$ is given as
\begin{eqnarray}
  \mathbf{F}
    &=&
  -2\int_0^\infty \mathrm{d}\omega\,\int \frac{\mathrm{d}^2\mathbf{k}}{(2\pi)^2}\nonumber \\
	  & &
	\times \mathbf{k}\, \mathrm{Tr}\left[
                                    \underline{S}(\mathbf{k}\cdot\mathbf{v}-\omega;\mathbf{v})
                                    \cdot
                                    \underline{G}_I(\mathbf{k},z_a,\omega)
                                 \right] \,.
  \label{eq:QFint}
\end{eqnarray}
Here, $\mathbf{k}$ is the component of the wavevector parallel to the surface, $\underline{G}$ 
is the Fourier transform with respect to the planar coordinates of the electromagnetic Green 
tensor. The tensor
\begin{eqnarray}
  \underline{S}(\omega) = \frac{1}{2\pi}\int_{-\infty}^\infty \mathrm{d}\tau\,e^{i\omega\tau}\,
                          \langle
                            \hat{\mathbf{d}}(\tau)\hat{\mathbf{d}}(0)
                          \rangle_\mathrm{NESS}\,
\end{eqnarray}
represents the power spectrum corresponding to the dipole-dipole correlator for the system's 
nonequilibrium steady state (NESS) and describes the strength of the fluctuations affecting 
the atomic system.
The subscripts ``$I$'' and ``$R$'' appearing in the previous and the following expressions 
denote the real and the imaginary part of the quantities they are appended to (e.g., 
$\underline{G}_I=\mathrm{Im}\lbrace \underline{G}\rbrace$, 
$\underline{G}_R=\mathrm{Re}\lbrace \underline{G}\rbrace$, etc.). 
Assuming that $\hat{q}(t)$ can be described in terms of a harmonic oscillator, the power 
spectrum can be written as \cite{Intravaia2016a}
\begin{equation}
  \underline{S}(\omega; \mathbf{v}) = \frac{\hbar}{\pi}\theta(\omega)\underline{\alpha}_I(\omega; \mathbf{v}) \,
	+ 
	\frac{\hbar}{\pi}\underline{J}(\omega; \mathbf{v}),
\label{PS}
\end{equation}
where $\underline{\alpha}(\omega; \mathbf{v})$ is the velocity-dependent dressed atomic
polarizability (see Appendix~\ref{app1}).
The first term on the r.h.s. corresponds to the result one would obtain using the so-called
local thermal equilibrium (LTE) approximation. Within this approximation, it is assumed that
the atom is in equilibrium with its immediate surroundings, allowing the application of the
fluctuation-dissipation theorem  \cite{Callen1951}. The ``locally equilibrated atom"' is
subsequently coupled to the substrate material.
However, a full nonequilibrium description yields the additional term 
$\underline{J}(\omega; \mathbf{v})$,
which substantially contributes to the quantum frictional process (see Appendix~\ref{app1}
and Refs. \cite{Intravaia14,Intravaia2016a,Reiche2017}).

The physics of quantum friction is connected to that of the quantum Cherenkov effect through 
the anomalous Doppler effect \cite{Pieplow15,Maghrebi13a,Intravaia2016}. 
In simple terms, we have that, through the Doppler shift appearing in Eq.~\eqref{eq:QFint}, 
this process brings negative frequencies of the electromagnetic spectrum into the integration
region which is physically relevant for the interaction.
Previous work has shown that, depending on the atom's velocity, quantum friction is characterized 
by the combination of a non-resonant and a resonant contribution. The resonant part occurs when 
the system's resonances, such as atomic transition frequencies or polaritonic surface modes 
existing at the vacuum/substrate interface, participate in the interaction. Usually, they become 
relevant only for velocities high enough to generate a Doppler shift, which displaces the resonances 
into the aforementioned relevant frequency range. As a rough rule of thumb, this occurs for 
$v/z_{a}>\omega_{\rm r}$, where $\omega_{\rm r}$ is the resonance frequency 
under consideration.   
Similarly, the non-resonant part gives the dominant contribution for the force at low velocities 
and is directly related to the low-frequency optical response of the substrate. Specifically, 
this region is  strongly affected by the dissipative behavior of the material(s) composing the
substrate. In this non-resonant regime the force is to a good approximation described by
\begin{widetext}
\begin{align}
   \mathbf{F}\approx 
	    & 
	  -2\frac{\hbar}{\pi} \int \frac{\mathrm{d}^{2}\mathbf{k}}{(2\pi)^{2}} \, \mathbf{k}\,\theta(\mathbf{k}\cdot\mathbf{v})
		                    \int \frac{\mathrm{d}^{2}\tilde{\mathbf{k}}}{(2\pi)^{2}}
												\int_{0}^{\mathbf{k} \cdot\mathbf{v}} \mathrm{d}\omega\,
                           \mathrm{Tr}\left[\underline{\alpha}_{0} \cdot \underline{\sigma}_{I}(\tilde{\mathbf{k}},z_{a},[\mathbf{k}+\tilde{\mathbf{k}}] \cdot\mathbf{v}-\omega)\right]
													 \mathrm{Tr}\left[\underline{\alpha}_{0}\cdot \underline{\sigma}_{I}(\mathbf{k},z_{a},\omega)\right]\nonumber\\
      &
	  -2\frac{\hbar}{\pi} \int \frac{\mathrm{d}^{2}\mathbf{k}}{(2\pi)^{2}} \, \mathbf{k}
		                    \int \frac{\mathrm{d}^{2}\tilde{\mathbf{k}}}{(2\pi)^{2}} \theta(\mathbf{\tilde{k}} \cdot \mathbf{v})
												\int_{\mathbf{k}\cdot\mathbf{v}}^{[\mathbf{k}+\tilde{\mathbf{k}}]\cdot\mathbf{v}}\mathrm{d}\omega \,
                           \mathrm{Tr}\left[\underline{\alpha}_{0}\cdot \underline{\sigma}_{I}(\tilde{\mathbf{k}},z_{a},[\tilde{\mathbf{k}}+\mathbf{k}]\cdot\mathbf{v}-\omega)            
													                  \cdot \underline{\alpha}_{0}
																						\cdot \underline{\sigma}_{I}(\mathbf{k},z_{a},\omega)
																		  \right].
  \label{Fsmallv}
\end{align}
\end{widetext}
Here, the dyadic $\underline{\alpha}_0 = \mathbf{d}\mathbf{d}$ describes the static polarizability
for our model. In the above expression we have also used that, due to the properties of the trace
and of the polarizability, we can replace the Green tensor by
its (symmetric) diagonal part, $\underline{\sigma} (\mathbf{k},z_a,\omega)$.
Since quantum friction strongly decays with increasing atom-surface separation (see also Sec.~\ref{Sec:QFres}),
the dominant contribution of the above expressions come from the system's near-field region. In
this region $\underline{\sigma} (\mathbf{k},z_a,\omega)$ can be written as
\begin{equation}
 \underline{\sigma}(\mathbf{k},z_a,\omega)
 \approx
 r^p(\omega,k)\mathrm{diag}\left[\frac{k_x^2}{k^2}, \frac{k_y^2}{k^2},1\right]\frac{k e^{-2k z_a}}{2\epsilon_0}
 \, .
\label{eq:greenenear}
\end{equation}
Here, $\epsilon_0$ is the vacuum permittivity, $k=|\mathbf{k}|$ and $r^p(\omega,k)$ is the reflection
coefficient of the substrate for the $p$-polarized electromagnetic radiation.

The previous equations show that the quantum frictional interaction is mainly connected with the 
$p$-polarized electromagnetic field (the $s$-polarized field gives a small contribution of the 
order $v^2/c^2$, with $c$ the speed of light) and is dominated by wave vectors $k\lesssim 1/z_{a}$ 
and frequencies $0<\omega\lesssim v/z_{a}$. 
It is interesting to note that, if in this regime we can write
$r^p_{I}(\omega,k)\approx 2 (\omega \epsilon_{0})^{n} \rho_{n}(k)$, Eq. \eqref{Fsmallv} gives a velocity dependence $F \propto v^{2n+1}$, while the distance
dependence is related to the detail of the generalized resistivity $\rho_{n}(k)$.
For $n=1$ one speaks of ohmic materials, while $n<1$ and $n>1$ indicate, respectively, sub-ohmic 
and super-ohmic behavior. This feature has been connected to the non-Markovian properties of the 
electromagnetic atom-surface interaction \cite{Intravaia14,Intravaia2016} and explains why many 
of the authors have obtained $F\propto v^3$ for the low-velocity asymptotic expression of the 
quantum frictional force on an atom moving above a substrate made of a homogeneous (ohmic) 
material \cite{Dedkov02a}.
Since the nature of the planar medium determines the functional dependence of the force, tailoring 
the properties of the substrate allows for a control of the interaction.

%%%%%%%%%%%%%%%%%%%%%%%%%%%%%%%%%

\section{Electromagnetic scattering near nano-structures}
\label{Sec:supEMA}

The above expressions highlight the dependence of the quantum frictional force on the optical 
response of the substrate and show the important role of the reflection coefficients of the substrate.
The literature offers many different approaches for calculating these 
quantities for nanostructures. However, most of the papers focus on frequency ranges, wave-vectors 
and, in general, material characteristics which are not those that are relevant for the evaluation 
of the quantum frictional force. In order to define the notation and give a consistent framework 
to our considerations, we present in this section an analysis which focuses on these aspects.

\begin{figure}[tb]
\resizebox{1.05\linewidth}{!}{
\includegraphics[scale=1]{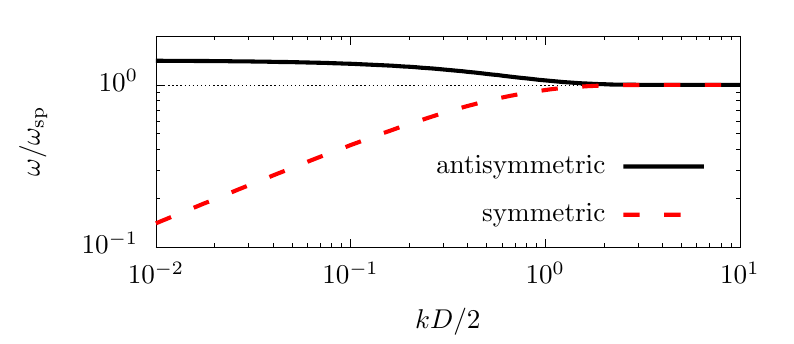}}
\caption{Dispersion relations of the symmetric and antisymmetric surface plasmon-polariton 
         modes of a slab of thickness $D$ consisting of a material described by a Drude model
         without dissipation, i.e. $\epsilon(\omega)=1-\omega_\mathrm{p}^2/\omega^2$. 
				 For $kD/2\gg 1$ both modes merge to the surface plasmon-polariton at $\omega_\mathrm{sp}$
				 for a half-space problem.
\label{Dispersion}}
\end{figure}

\begin{figure}[tbp]
\resizebox{1.05\linewidth}{!}{
\includegraphics[scale=1]{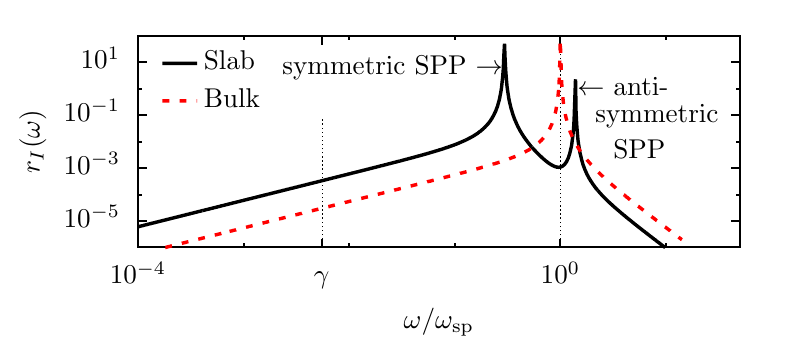}}
\caption{Frequency dependence of the imaginary part of the reflection coefficient 
         in the near-field limit. Two geometries with the same Drude material 
				 [Eq. \eqref{eq:dielfun}] are considered: A semi-infinite bulk (dotted 
				 red line) and a finite slab (solid black line). As parameters we chose 
				 typical values for gold \citep{Barchiesi2014} 
				 $\omega_\mathrm{p}=9\,\mathrm{eV}$ and $\gamma=35\,\mathrm{meV}$, 
				 $\epsilon_\mathrm{A}^\infty=1$, a slab thickness of 
				 $D=2\,\mathrm{nm}\approx10^{-3}\,c/\omega_\mathrm{p}$ and 
				 $k=\omega_\mathrm{p}/c$. 
				 For the half-space case, a resonance appears at the frequency 
				 $\omega_{\rm sp} \approx\omega_{\rm p}/\sqrt{2}$, while for the slab 
				 two resonances are visible, one above, the other below 
				 $\omega_{\rm sp}$.
				 \label{fig:refDL}
}
\end{figure}

We start by considering the expression for the reflection coefficients of a flat surface. In general, 
they can be written as \cite{Ford84}
\begin{equation}
  r^{\sigma}(\omega,k)
	  = 
  \delta^{\sigma} \frac{Z_0^{\sigma}(\omega,k)-Z_{\mathrm{m}}^{\sigma}(\omega,k)}{Z_0^{\sigma}(\omega,k)+Z_{\mathrm{m}}^{\sigma}(\omega,k)},
\label{eq:refZ}
\end{equation}
where the index $\sigma=s,p$ denotes the polarization state of light and we have introduced 
$\delta^{s/p}=\mp$. Further $Z_{\mathrm{m}}^{\sigma}(\omega,k)$ and $Z_{0}^{\sigma}(\omega,k)$
denote, respectively, the surface impedance for the substrate material and the material surrounding 
it (for simplicity, in the subsequent discussions, we assume this material to be 
vacuum).
The surface reflection coefficients are sensitive to the substrate material properties and 
the geometry of the system. For our forthcoming analyses, it is interesting to consider first 
the case of a slab of thickness $D$ suspended in vacuum and made by a homogeneous material characterized by the 
\emph{spatially local} complex permittivity function $\epsilon(\omega)$ \cite{Note1}.
%
%\footnote{In order to focus on the nano-structuring, we neglect the influence of \textit{non-local} effects in material properties throughout all the paper. For a detailed study of non-locality in this context of quantum friction see Ref.\cite{Reiche2017}.}. 
%
In this case Eq. \eqref{eq:refZ} simplifies as follow \cite{Sipe1981}
\begin{equation}
  r_\mathrm{slab}^\sigma(\omega,k) = r^\sigma_\mathrm{bulk}(\omega,k) \frac{1-e^{2ik_zD}}{1-[r^\sigma_\mathrm{Bulk}(\omega,k) e^{ik_zD}]^2},
\label{eq:refslab}
\end{equation}
where $k_z=\sqrt{\epsilon(\omega)\frac{\omega^2}{c^2}-k^2}$ ($\mathrm{Im}\lbrace k_z \rbrace \geq 0$,
      $\mathrm{Re}\lbrace k_z \rbrace \geq 0$), and $r^\sigma_\mathrm{Bulk}(\omega,k)$ is the 
interface reflection coefficient given by the usual Fresnel expressions \cite{Jackson75}.
For a spatially local, isotropic and homogenous substrate material we have that the impedance 
in Eq.~\eqref{eq:refZ} can be written as 
\begin{subequations}
\begin{gather}
  Z_\mathrm{m}^{s}(\omega,k)\equiv Z_\mathrm{bulk}^{s}(\omega,k) 
	  = 
  \frac{\frac{\omega^{2}}{c^{2}}}{\sqrt{\frac{\omega^{2}}{c^{2}}\epsilon(\omega)-k^{2}}}, \\
  Z_\mathrm{m}^{p}(\omega,k)\equiv Z_\mathrm{bulk}^{p}(\omega,k) 
	  =
	\frac{\sqrt{\frac{\omega^{2}}{c^{2}}\epsilon(\omega)-k^{2}}}{\frac{\omega^{2}}{c^{2}}\epsilon(\omega)},
\end{gather} 
\end{subequations}
while $Z_{0}^{\sigma}(\omega,k)$ can be obtained for our vacuum by setting $\epsilon(\omega) \equiv 1$.
The exponential in Eq.~\eqref{eq:refslab} represents the phase that is accumulated via the 
propagation and decay through the slab. 
The coefficient $r_\mathrm{slab}^\sigma(\omega,k)$ is characterized by two resonances, physically 
related to the interaction between the surface polaritons existing on either side of the slab 
\cite{Camley1984,Berini00}. 
This is best seen in the near-field region, where just one of the two polarizations contributes 
to the scattering process and the reflection coefficients $r^\sigma_\mathrm{Bulk}(\omega,k)$ take 
on the form
\begin{equation}\label{eq:bulkref}
 r^p_\mathrm{bulk}(\omega,k) \approx \frac{\epsilon(\omega)-1}{\epsilon(\omega) +1}\,,\quad
 r^s_\mathrm{bulk}(\omega,k) \approx 0\,.
\end{equation}
The dispersion relations of the above-mentioned polaritonic modes are given by the solutions 
of (see also Fig. \ref{Dispersion})
\begin{equation}
  \epsilon(\omega)=
  -\begin{cases}
     \coth\left(\frac{k D}{2}\right)\quad \to\, \mathrm{symmetric}\\
     \tanh\left(\frac{k D}{2}\right)\quad \to\, \mathrm{antisymmetric}
   \end{cases}.
\label{eq:slabdisp}
\end{equation}
The two coupled surface polaritons are labeled symmetric and antisymmetric in relation to 
the properties of the electric fields they are associated with. Due to the different field 
distributions within the slab, the symmetric polariton has a lower energy (or frequency) 
than antisymmetric, with the uncoupled surface polariton's energy lying in between both. 
The splitting between of the symmetric and the antisymmetric surface polariton increases 
as $1/kD$ and it is, therefore, more pronounced for thin slabs, for which the coupling 
between the surface excitations is stronger. 
If different dielectric materials were used below and above the slab, additional leaky 
modes would come into play as elaborated in Ref.~\cite{Burke1986}. The solutions of 
Eq.~\eqref{eq:slabdisp} are clearly visible in the imaginary part of the reflection coefficient 
as shown in Fig.~\ref{fig:refDL}, where they are also compared to the resonance of a 
semi-infinite homogeneous substrate. In Fig.~\ref{fig:refDL} we consider a metal described by the 
Drude model
\begin{equation}
  \epsilon(\omega) = \epsilon^\infty- \frac{\omega_\mathrm{p}^2}{\omega(\omega+i\gamma)},
\label{eq:dielfun}
\end{equation}
where $\epsilon^\infty>0$ describes the response of the material at large frequencies, 
$\gamma$ denotes a phenomenological damping constant and $\omega_\mathrm{p}$ the plasma 
frequency. In this case the resonances in the reflection coefficient are associated with 
the so-called surface plasmon-polaritons. For a bulk-vacuum interface, in the near-field 
limit, the resonance is located at $\omega_{\rm sp}=\omega_{\rm p}/\sqrt{1+\epsilon^\infty}$, 
while in the case of the slab they depend on the wave-vector and both tend to 
$\omega_{\rm sp}$ for $kD\to \infty$.

Notice that, in Fig.~\ref{fig:refDL} the behavior at low frequencies (i.e., frequencies 
much smaller than the resonance frequency) is similar for both the bulk and the slab and 
describe an ohmic response of both structures. Indeed, in this region, assuming that the 
material composing the slab or the bulk is ohmic, an expansion of the imaginary part of the reflection 
coefficient gives 
\begin{equation}
  r_{I}^p(\omega,k)
	  \stackrel{\omega\ll\omega_\mathrm{sp}}{\approx}
  \omega\epsilon_{0}
    \begin{cases}
      2 \rho & \text{for~bulk}\\
      2 \rho  \coth[D k] & \text{for~slab}
    \end{cases}\,.
\label{eq:taylorslab}
\end{equation}
Here, $\rho$ represents the material resistivity ($\rho=\gamma/(\epsilon_0\omega_\mathrm{p}^2)$ 
for the Drude model). Notice that, since $k>0$, the imaginary part of the reflection 
coefficient at low frequencies increases with thinner slabs. 
In addition, we would have obtained the same result even if the ohmic layer were deposited above a dielectric bulk instead of being suspended in vacuum.
These results can be understood in relation to the behavior the symmetric polaritonic resonance, which in case 
of metals is sometimes called short-range plasmon-polaritons \cite{Berini00,Berini09}. 
The field corresponding to the symmetric mode is indeed more confined within the slab
material and thus exhibits a stronger dissipative response than both, the single-interface 
resonance (bulk reflection coefficient) and the anti-symmetric mode (which is sometimes
also referred to as the long-range plasmon-polaritons \cite{Berini00,Berini09}).
For a superlattice structure made by a semi-infinite stack of alternating layers of two 
different materials (labeled A and B hereafter) with respective thickness $d_{\rm A}$ 
and $d_{\rm B}$ \cite{Albuquerque2004}, the expressions for the reflection coefficients 
become more involved.
To calculate them we need to replace the surface impedance in Eq.~\eqref{eq:refZ} with 
that of the superlattice, $ Z_\mathrm{sup}^\sigma(\omega,k)$. This can be calculated through the transfer matrix formalism 
\cite{Yariv1984}. Within this approach, one propagates the electromagnetic field through 
each layer and fulfills the boundary conditions at each interface. For instance, the 
propagation through the layer A is given by
\begin{equation}
  \begin{pmatrix}
    \mathbf{E} \\ c\,\mathbf{B}
  \end{pmatrix}_{z=z_0^-+d_{\rm A}}^\sigma
  =
  \mathbb{M}_{\rm A}^\sigma(d_{\rm A})\,
  \begin{pmatrix}
    \mathbf{E} \\ c\,\mathbf{B}
  \end{pmatrix}_{z=z_0^-}^\sigma\,,
\end{equation}
where $z_0^-$ indicates the position directly in front of the interface. The transfer 
matrix through the local material A reads
\begin{equation}
  \mathbb{M}^\sigma_\mathrm{A}(d_{\rm A}) = 
  \begin{pmatrix}
    \cos(k_z^\mathrm{A}\, d_{\rm A})                                     & i\delta^\sigma\sin(k_z^\mathrm{A}\, d_{\rm A}) \, Z^\sigma_\mathrm{A} \\
    i\delta^\sigma\sin(k_z^\mathrm{A}\, d_{\rm A}) / Z^\sigma_\mathrm{A} & \cos(k_z^\mathrm{A}\, d_{\rm A})
  \end{pmatrix}\,.
\end{equation} 
For non-local materials the transfer matrix takes a different form and explicit 
expressions can be found in Ref. \cite{Intravaia2015}.
If we stack the layers A and B we can describe the propagation through the 
combined block of thickness $d_{\rm sl}=d_{\rm A}+d_{\rm B}$ with the transfer 
matrix
$\mathbb{T}=\mathbb{M}_\mathrm{B}\mathbb{M}_\mathrm{A}$ or $=\mathbb{M}_\mathrm{A}\mathbb{M}_\mathrm{B}$, 
depending on the stacking sequence \cite{Kidwai12}. 
Using the Bloch theorem  \cite{Bloch1929} for periodic structures, we 
obtain \cite{Mochan1987} 
\begin{equation}
  Z_\mathrm{sup}^\sigma(\omega,k)
  =
	\frac{\mathbb{T}^\sigma_{12}}{\exp(i\beta^\sigma d_{\rm sl})-\mathbb{T}^\sigma_{11}}
  =
	\frac{\exp(i\beta^\sigma d_{\rm sl})-\mathbb{T}^\sigma_{22}}{\mathbb{T}^\sigma_{21}}\,.
\end{equation}
The Bloch wavevector $\beta^\sigma$ can be related to the other parameters of the system 
through the  implicit dispersion relation
\begin{eqnarray}
 \cos(\beta d_{\rm sl})
 & = &
 \cos(k_z^\mathrm{A} d_\mathrm{A})\cos(k_z^\mathrm{B} d_\mathrm{B})
 -\frac{1}{2}\left(\frac{\epsilon_\mathrm{A}k_z^\mathrm{B}}{\epsilon_\mathrm{B}k_z^\mathrm{A}}
                   +
                   \frac{\epsilon_\mathrm{B}k_z^\mathrm{A}}{\epsilon_\mathrm{A}k_z^\mathrm{B}}
             \right)\nonumber \\
 &   & 
 \times \sin(k_z^\mathrm{A} d_\mathrm{A})\sin(k_z^\mathrm{B} d_\mathrm{B})\,.
\end{eqnarray}
Since most of our considerations will address the $p$-polarization (see the discussion
above), we drop hereafter the superscript (analogous  expressions hold for the $s$-polarization).
In addition, we focus on systems composed of alternating conducting and dielectric layers, 
where the stacking sequence starts with a conducting layer.

Similarly to the coupled surface polaritons found in the slab, an ensemble of excitations linked 
to the interaction among all the interface modes of the stacking sequences appears in the 
superlattice system. We refer to this ensemble as collective interface plasmon-polaritons (CIPPs) 
and their electromagnetic behavior at the vacuum-superlattice interface is, to some extent, similar 
to that of bulk plasmons occurring in the nonlocal description of metals \cite{Camley1984,Barton79,Note2}.
%
%\footnote{In addition to the CIPP, when the thickness of the metallic layer is larger than that of the dielectric layer ($d_{\rm A}>d_{\rm B}$), some other modes can appear in the electromagnetic spectrum characterizing the system \cite{Camley1984}. For simplicity we will exclude them from the present investigation, by limiting our analysis to the case $d_{\rm A}\le d_{\rm B}$.}.
%

\begin{figure}
\resizebox{1.1\linewidth}{!}{
\includegraphics[scale=1.]{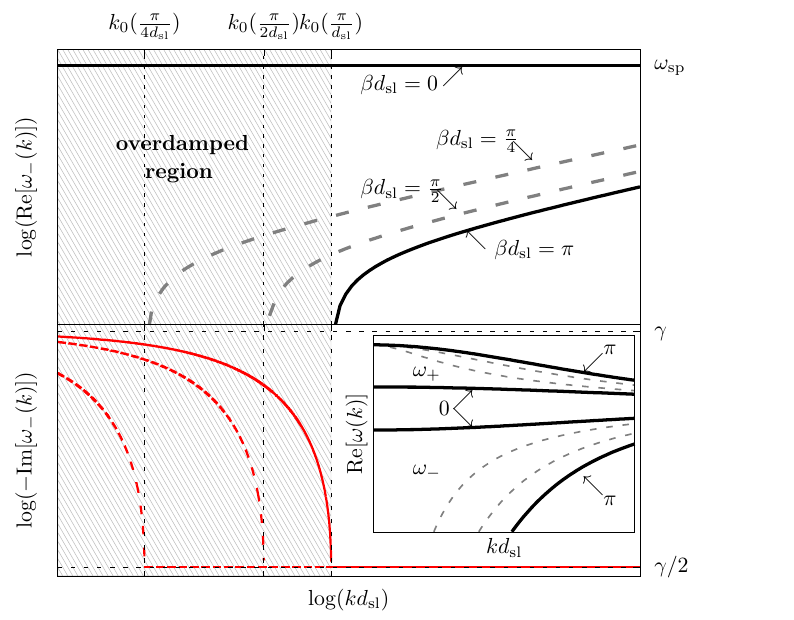}}
\caption{
         \textit{Top panel}: Dispersion relation of the $\omega_-$ branch on a double logarithmic scale. 
				 The thick black solid lines mark the edges of the branch with $\beta d_\mathrm{sl}=0,\,\pi$. 
				 Some intermediate values for the Bloch vector are also represented (dashed gray lines).
         The shaded area refers to a region, where the modes become overdamped. For different values
				 of $\beta$, the value of $k_0(\beta)$, below which the modes becomes overdamped, varies and 
				 shifts to lower $k$ for higher $\beta$.
         \textit{Bottom panel}: Negative imaginary part of the $\omega_-$ modes. A clearly visible 
				 jump occurs for $k\sim k_0(\beta)$ and the damping increases from $\gamma/2$ to $\gamma$. 
				 \textit{Inset}: Dispersion relation of the $\omega_\pm$ branches, again with the different 
				 lines referring to different values of $\beta$, in analogy to the top panel.
\label{fig:suplatbranchlog}
}
\end{figure}

Indeed, nanostructuring adds to the optical response of the medium certain features which are 
mathematically reminiscent of spatial nonlocality, although the individual constituents are 
described in terms of a spatially local permittivity \cite{chebykin11}. However, in contrast
to bulk plasmons in nonlocal metals, the CIPP fields inside the nanostructured materials are 
always transverse. In the near-field approximation their dispersion relations are solutions 
of (see Ref.~\cite{Camley1984})
\begin{equation}
  \frac{\epsilon_\mathrm{A}(\omega)}{\epsilon_\mathrm{B}(\omega)}
	=
	-C(k,\beta)\pm\sqrt{C^2(k,\beta)-1},
\label{eq:disprel}
\end{equation}
where
\begin{equation}
  C(k,\beta) 
	= 
  \frac{\cosh(kd_\mathrm{A})\cosh(kd_\mathrm{B})-\cos(\beta d_\mathrm{sl})}{\sinh(kd_\mathrm{A})\sinh(kd_\mathrm{B})}\,.
\label{eq:c}
\end{equation}
Adopting the notation used in Ref.~\cite{Camley1984}, we write $C(k,\beta)\mp\sqrt{C^2(k,\beta)-1}=\exp[\pm\psi(k,\beta)]$.
Upon using the Drude model as in Eq.~\eqref{eq:dielfun} for material A (metal) and a dielectric 
constant for material B \cite{Note3}, the explicit dispersion relation reads
\begin{equation}
  \omega_\pm(k,\beta)
  = 
	-\frac{i\gamma}{2} +\sqrt{\frac{\omega_\mathrm{p}^2}{\epsilon_\mathrm{A}^\infty+\epsilon_\mathrm{B} \exp[\mp\psi(k,\beta)]}
                            -\frac{\gamma^2}{4}}\,.
\label{eq:omegapm}
\end{equation}
The $\omega_\pm$ denote two different branches of possible solutions of Eq.~\eqref{eq:disprel}. Similar 
to the result of Eq.~\eqref{eq:slabdisp} for slabs, the two branches can be associated with symmetric 
($\omega_-$) and antisymmetric ($\omega_+$) modes. In fact, depending on the number of supercells, 
a finite superlattice structure exhibits many distinct symmetric and antisymmetric modes parametrized 
by discrete values of the Bloch vector \cite{Johnson1985}.
When the periodic pattern is repeated an infinite number of times, the distinct lines of a finite superlattice structure 
blur into a continuum \cite{Camley1984} (see Fig.~\ref{fig:suplatbranchlog}).
Within such a limit, the real part of the Bloch vector $\beta_R$ continuously varies within the Brillouin 
zone $[0,\pi/d_{\rm sl}]$, while the imaginary part $\beta_I$ has to be positive in order to obtain a 
decaying field away from the surface. For non-dissipative material, as a function of the Bloch vector 
each branch spans two areas on the $(k,\omega)$-plane, which characterize the continua of the symmetric 
and antisymmetric modes. These areas are bounded by the curves obtained from Eq.~\eqref{eq:omegapm} for 
$\beta= 0$ and $ \beta=\pi/d_{\rm sl}$.
Due to damping within the metallic material some of the low frequency modes belonging to the symmetric 
$\omega_-$ branch become overdamped for small wave-vectors. 
This occurs for the $\omega_-$ branch for ($kd_\mathrm{sl}\ll1 $; see Fig.~\ref{fig:suplatbranchlog})
\begin{equation}
  k < k_\mathrm{0}(\beta)=\frac{\gamma}{\omega_\mathrm{p}}\sqrt{\epsilon_\mathrm{B}\frac{1-\cos(\beta d_{\rm sl})}{2d_\mathrm{A}d_\mathrm{B}}}~.
\label{eq:overdamp}
\end{equation}
In this overdamped region, the modes exhibit a purely imaginary frequency. For example, the lower boundary 
of the $\omega_-$ branch obtained for $\beta=\pi/d_{\rm sl}$ tends to $-\imath\gamma$ for $k< k_\mathrm{0}(\pi/d_{\rm sl})$. 
For $\beta=0$, the frequencies is pure imaginary only for $k=0$, indicating that the frequency of the 
modes near the upper bound of the $\omega_-$ branch and the lower bound of the $\omega_+$ branch have 
a nonvashing real part for all wave vectors.

Composite nanostructures such as those discussed above are often described through the so-called effective medium approximation (EMA) 
\cite{Bruggeman1935}. This approach relies on the fact that for wavelengths larger than the characteristic 
geometric length scale of the system (in our case, the thickness of the supercell $d_\mathrm{sl}$), 
the electromagnetic field cannot resolve the details of the system. Instead, the electromagnetic field
effectively averages the structural details so that the nanostructures can be described through an
effective dielectric function. This approach drastically simplifies the description of complex nanostructures, 
revealing features which are often obscured by an involved mathematical machinery. Depending on the 
geometry of the composite system, the resulting effective dielectric function may indeed exhibit 
properties that are different from those of the constitutive elements. For instance, for our superlattice 
structures, the EMA describes the system as an uniaxial crystal with a dielectric tensor 
$\underline{\epsilon}_{\rm EMA}(\omega) = \mathrm{diag}[\epsilon_\perp (\omega),\epsilon_\perp (\omega),\epsilon_\parallel(\omega)]$ 
whose entries are given by \cite{Mochan1987,Poddubny13}
\begin{subequations}
  \begin{gather}
    \epsilon_\perp (\omega)
		  =
    \epsilon_\mathrm{A}(\omega)\, f + \epsilon_\mathrm{B}(\omega)\, (1-f),\\
    \epsilon_\parallel(\omega)
		  =
		\left[\frac{f}{\epsilon_\mathrm{A}(\omega)} + \frac{1-f}{\epsilon_\mathrm{B}(\omega)}\right]^{-1},
\end{gather}
\label{epsilonEMA}
\end{subequations}
where $f=d_\mathrm{A}/d_{\rm sl}$ gives the filling factor of material A. Equations \eqref{epsilonEMA} 
correspond to the propagation of the electromagnetic field parallel ($\epsilon_\parallel$) or orthogonal ($\epsilon_\perp$) to the optical axis 
of the crystal, in our case the $z$-axis. Compared to an isotropic bulk material, in an uniaxial crystal 
the dielectric response differs along the different principal axes. Specifically, in uniaxial crystals 
with the optical axis perpendicular to the surface, ordinary waves are associated with the $s$-polarization, 
whereas extraordinary waves are associated with the $p$-polarization \cite{Landau84}. For such systems, 
the surface impedances are sensitive to the anisotropy and are given by \cite{Knoesen1985}
\begin{subequations}
  \begin{gather}
    Z_\mathrm{EMA}^{s}(\omega,k)
		 =
		\frac{\frac{\omega^{2}}{c^{2}}}{\sqrt{\frac{\omega^{2}}{c^{2}}\epsilon_\perp (\omega)-k^{2}}}\\
    Z_\mathrm{EMA}^{p}(\omega,k)
		 =
		\frac{\sqrt{\frac{\omega^{2}}{c^{2}}\epsilon_\parallel(\omega)-k^{2}}}{\frac{\omega^{2}}{c^{2}}\sqrt{\epsilon_\perp(\omega)\epsilon_\parallel(\omega)}}.
\end{gather} 
\end{subequations}
In the near-field, the corresponding reflection coefficients take on the form 
\begin{equation}
  r_\mathrm{EMA}^p(\omega,k)
	  \approx
  \frac{\epsilon_\mathrm{eff}(\omega)-1}{\epsilon_\mathrm{eff}(\omega)+1}\,,\quad
  r_\mathrm{EMA}^s(\omega,k)
	  \approx 
  0,
\label{eq:refEMA}
\end{equation}
where we have introduced the effective dielectric function as the geometric mean of the
perpendicular and parallel components of the dielectric according to
$\epsilon_\mathrm{eff}(\omega) = \sqrt{\epsilon_\parallel(\omega)\epsilon_\perp(\omega)}$. 

If we now consider the case where $|\epsilon_\mathrm{A}(\omega)|\gg|\epsilon_\mathrm{B}(\omega)|$ 
we can, for a certain frequency range, reduce this effective dielectric function to 
\begin{equation}
  \epsilon_\mathrm{eff} (\omega)
	  \approx
  \sqrt{\frac{f}{1-f}\epsilon_\mathrm{A} (\omega) \epsilon_\mathrm{B} (\omega)}
	  =
	\sqrt{\frac{d_{\rm A}}{d_{\rm B}}\epsilon_\mathrm{A} (\omega) \epsilon_\mathrm{B} (\omega)}\,.
\label{eq:epseff}
\end{equation} 
If in this limit $\epsilon_\mathrm{B}>0$ is a constant \cite{Note3}, then $\epsilon_\mathrm{eff}(\omega) \propto \sqrt{\epsilon_\mathrm{A}(\omega)}$.
In essence, this yields a criterion when the EMA provides a significant deviation from 
the ordinary optical response of a bulk system made purely by the material A. To see this 
more clearly, consider, as an example, an ohmic material which at low frequencies behaves 
as $\epsilon_\mathrm{A}(\omega)\approx \imath(\omega \epsilon_{0}\rho)^{-1}$ (e.g., a 
Drude metal for $\omega<\gamma$). Within the effective medium description, for the reflection coefficient
 we then have 
\begin{equation}
  r_{I}(\omega,k)
	  \approx 
	\sqrt{\omega\epsilon_{0}} \sqrt{2\frac{\rho}{\epsilon_\mathrm{B}}\frac{d_{\rm B}}{d_{\rm A}}}.
\label{rIsuplatt}
\end{equation}
Therefore in the case of a metal-dielectric superlattice structure, the EMA predicts that 
for $\omega<\gamma$ the behavior of the reflection coefficient is no longer ohmic but sub-ohmic.

\begin{figure}
\resizebox{1.05\linewidth}{!}{
\includegraphics[scale=1]{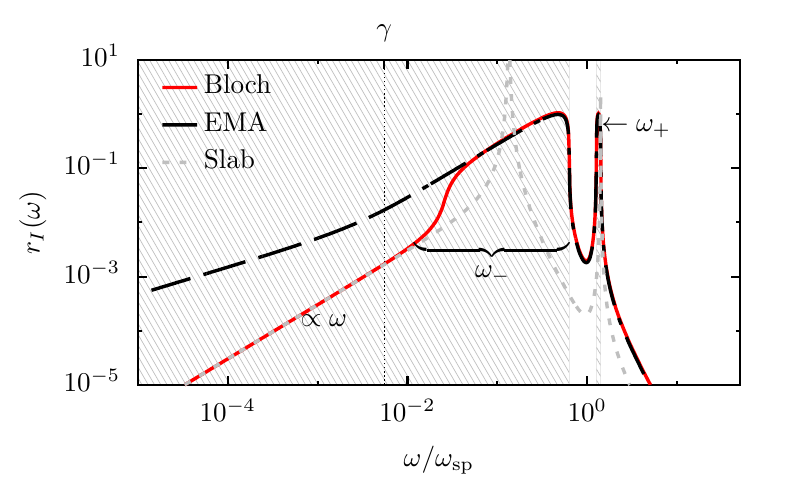}}
\caption{Limiting behavior of the superlattice's reflection coefficient ($p$-polarization) 
         versus the full calculation. The imaginary part of the reflection coefficient is 
				 plotted as a function of $\omega$. The conducting material's dielectric function, 
				 $\epsilon_{\rm A}(\omega)$, is described by a Drude model with the same parameters 
				 as in Fig.~\ref{fig:refDL}. The dielectric is vacuum ($\epsilon_{\rm B}=1$), while 
				 the filling factor and the wave-vector are set to $f=0.2$ and $k=10^{-1}c/\omega_\mathrm{p}$. 
 				 The full tranfer-matrix-based calculation exhibits features which can be directly 
				 connected to the continuum of modes in the $\omega_{\pm}$-branches. For frequencies 
				 $\omega$ higher than the lower bound of the $\omega_{-}$-branch, the EMA calculation 
				 (dashed black line) shows a very good agreement with the transfer-matrix approach 
				 (solid red line). At small frequencies $\omega$, the full transfer-matrix calculation 
				 is equivalent to the behavior of the very first conducting layer of the structure 
				 (gray dotted line). The shaded areas represent the hyperbolic regime with
				 $\mathrm{Re}\left\lbrace\epsilon_\perp(\omega)\right\rbrace
				  \mathrm{Re}\left\lbrace\epsilon_\parallel(\omega)\right\rbrace < 0$.
				 \label{fig:refbreak}
}
\resizebox{1.05\linewidth}{!}{
\includegraphics[scale=1]{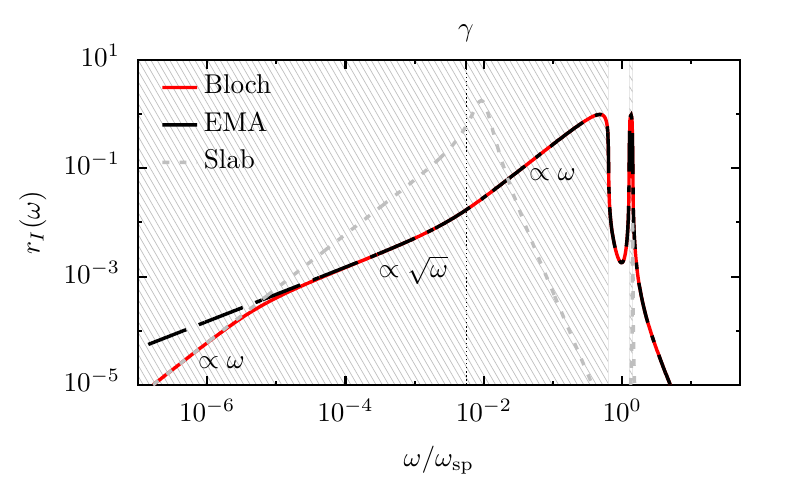}}
\caption{Analogous comparison as in Fig.~\ref{fig:refbreak} but this time for $k=10^{-3}c/\omega_\mathrm{p}$. 
         The EMA holds over a larger frequency range and for lower frequencies $\omega$. 
				 A sub-ohmic behavior ($r_\mathrm{I}\propto \sqrt{\omega}$) is visible in the 
				 Bloch-wave calculations. At low frequencies, the EMA breaks down and effectively 
				 only the first slab is responsible for the scattering properties of the entire
				 structure.
				 \label{fig:refsqrt}
}
\end{figure}

Figures~\ref{fig:refbreak} and \ref{fig:refsqrt} display the above features and certain 
structures related to the CIPP modes. Using the different approaches described above 
(Bloch waves and EMA), both plots display the frequency dependence $r^{p}_{I}(\omega,k)$ 
for two distinct values of the wave-vector, one above and one below the value $k_\mathrm{0}(\pi/d_{\rm sl})$
that delineates the overdamped region. 
Notice that the EMA agrees with the full calculation only above a certain frequency. 
The breakdown of the approximation occurs for frequencies around the lower boundary of 
the $\omega_-$ branch. This can be understood by recalling that in this region 
$\beta d_\mathrm{sl} \approx\pi$ (see Fig. \ref{fig:suplatbranchlog}), while previous 
work \cite{Mochan1987} has shown that the expressions in Eqs. \eqref{epsilonEMA} are 
only compatible with $\beta d_\mathrm{sl}\ll 1$. In both plots, we can see that for 
$\omega<\gamma$ the EMA description enters the sub-ohmic regime discussed above. This 
behavior is also featured by the full calculation as long as $\gamma$ lies above the 
lower boundary of the $\omega_-$ branch. 
Indeed, in Fig.~\ref{fig:refsqrt}, due to the choice of the wave-vector, the $\omega_-$ 
branch is stretched to lower frequencies $\omega$ and the lower bound is not marked 
by a distinct edge as that appearing in Fig.~\ref{fig:refbreak}. 
The sub-ohmic feature of the superlattice occurs in the region where the modes of the 
$\omega_-$ branch becomes overdamped, connecting it to the collective low frequency 
behavior of the (non-resonant) CIPP. Conversely, the shoulder appearing in 
Fig.~\ref{fig:refbreak} can be interpreted as resulting from the coalescence of all 
the (infinite) CIPP resonances occurring in the semi-infinite superlattice. 

The EMA also provides the framework for another interesting aspect of
superlattice structures (or in general uniaxial crystals), namely the appearance of hyperbolic 
dispersions \cite{Poddubny13}.
%\new{
%If we consider an $p$ polarized wave in an uniaxial medium with an optical
%axes perpendicular to the surface we find
%\begin{equation}
%\frac{k_x^2+k_y^2}{\epsilon_\perp(\omega)} +
%\frac{k_z^2}{\epsilon_\parallel(\omega)} = \frac{\omega^2}{c^2}\,.
%\label{eq:extrauni}
%\end{equation} 
%
%Assuming for a moment that $\epsilon_\perp$ and $\epsilon_\parallel$ are real constants,
%we can extract simple isofrequency contours.
Indeed, depending on the sign of $\mathrm{Re}\lbrace\epsilon_{\parallel,\perp}\rbrace$,
isofrequency surfaces in the 3D-wavevector space can be either ellipsoids or hyperboloids.
The latter occurs when $\mathrm{Re}\lbrace\epsilon_{\parallel}\rbrace\mathrm{Re}\lbrace\epsilon_{\perp}\rbrace<0$
and, depending on which of the permittivities is negative, one distinguishes between hyperbolic 
material of type I ($\mathrm{Re}\lbrace\epsilon_{\parallel}\rbrace<0$ and $\mathrm{Re}\lbrace\epsilon_{\perp}\rbrace>0$)
or of type II ($\mathrm{Re}\lbrace\epsilon_{\parallel}\rbrace>0$ and $\mathrm{Re}\lbrace\epsilon_{\perp}\rbrace<0$).
Distinct from a usual dispersion, in hyperbolic materials a large number of wavevectors can be connected 
with a narrow range of frequencies leading to a significant increase in the system's density of states \cite{Poddubny13}.
%Of course these clear shapes will be
%altered if we have dispersive and dissipative materials, but nevertheless we can assign the
%naming convention of \textit{hyperbolic} to regimes with
%$\mathrm{Re}\lbrace\epsilon_\perp(\omega)\rbrace \mathrm{Re}\lbrace\epsilon_\parallel(\omega)\rbrace < 0$.
In Fig. \ref{fig:refbreak} and \ref{fig:refsqrt} the shaded areas indicate where our superlattice behaves
as a hyperbolic material.
For a metal, modeled by a Drude model, with low damping ($\gamma\ll \omega_\mathrm{p}$) and a dielectric with constant $\epsilon_{\rm B}>0$, the frequencies where the relative sign flips are given by
\begin{subequations}
\begin{equation}
\omega_\mathrm{h1} \sim \omega_\mathrm{p}
\sqrt{\frac{f}{f\epsilon_\mathrm{A}^\infty + (1-f)\epsilon_\mathrm{B}}},
\end{equation}
\begin{equation}
\omega_\mathrm{h2} \sim \omega_\mathrm{p}
\sqrt{\frac{1-f}{f\epsilon_\mathrm{B} + (1-f)\epsilon_\mathrm{A}^\infty}},\quad
\omega_\mathrm{h3} \sim \frac{\omega_\mathrm{p}}{\sqrt{\epsilon_\mathrm{A}^\infty}}\,.
\end{equation}
\end{subequations}
We notice that, under the condition of validity of the EMA, this behavior is essentially related with the
location of the $\omega_{\pm}$-branches, establishing a direct connection with the CIPP \cite{Isic2017}. Interestingly, this offers another perspective on the features we observe in $r_{I}(\omega)$.
Indeed, if we exclude the sub-ohmic region, where the material dissipation is relevant, the shoulders appearing in the plots (in particular in Fig. \ref{fig:refbreak}) can be seen as a manifestation of the hyperbolic behavior of the semi-infinite superlattice. In fact, due to the change in sign of the permittivities, in this region $\mathrm{Im}\lbrace\epsilon_\mathrm{eff}(\omega) \rbrace$ and therefore $r_{I}(\omega)$ can be substantially different from zero even for a vanishingly small material damping.
This additional loss channel can be understood by the deep penetration of the CIPPs, which allows to accumulate even very small
losses throughout the whole semi-infinite superlattice substrate.

The above plots also highlight the relevance of the wavevector regarding the validity 
of the EMA, showing that the smaller the value of $k$ (near to orthogonal incidence)
becomes, the better is the quality of the EMA. 
Importantly, both Figs.~\ref{fig:refbreak} and \ref{fig:refsqrt} reveal that at low 
frequencies, below the area described by the $\omega_{-}$ branch of the CIPP, the 
EMA description ceases to be valid. In this case, the optical response of the superlattice 
structure essentially reduces to that of the first metallic layer in the system, recovering 
the ohmic behavior of a single slab. Physically, this can be understood as the result 
of a shorter penetration of the field into the structure: The EMA breaks down for 
penetration depths which are shorter than the thickness of the supercell (the field 
is no longer able to resolve deeper lying layers).

\begin{figure}
\resizebox{1.05\linewidth}{!}{
\includegraphics[scale=1]{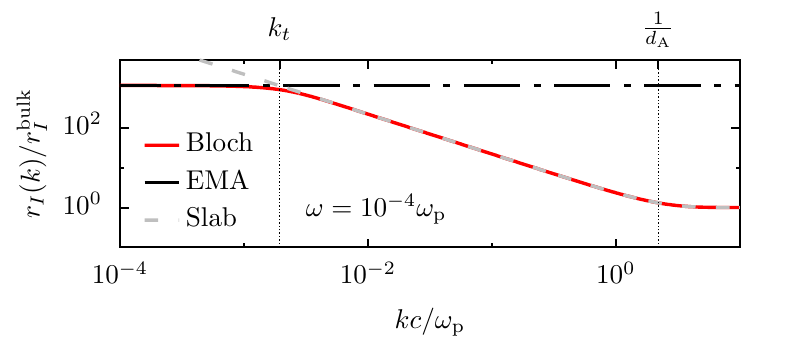}}
\caption{Wavevector dependence of the imaginary part of the $p$-polarized reflection coefficient 
         for $\omega=\sqrt{2}\times10^{-4}\omega_{\rm sp}$. 
         The material parameters were chosen analogous to Fig. \ref{fig:refDL} with 
				 $d_\mathrm{A}=d_\mathrm{B}=10\,\mathrm{nm}$ ($\sim 6\times 10^{-3} \, c/\omega_\mathrm{p}$).
         The results for the full Bloch-wave calculation (red line), the EMA (black dash-dotted 
				 line) and the slab description (dashed gray line) are normalized by the imaginary part 
				 of the reflection coefficient for a metallic half-space. The full calculation is well 
				 described by the EMA at low $k$ (near orthogonal incidence) and it recovers the slab 
				 and the bulk results for $k>k_{t}$ and $k>1/d_{\rm A}$, respectively. 
				 \label{fig:refkp}
}
\end{figure}

This is also visible in Fig.~\ref{fig:refkp} where the wavevector dependence of the
reflection coefficient $r^{p}_{I}(\omega,k)$ is depicted for a fixed frequency: For 
small wavevectors, the full result is indeed well represented by the EMA, while for 
large wavevectors we recover the slab's reflection coefficient. When this occurs, the 
transition between the ohmic and the sub-ohmic behavior is characterized by the wavevector 
$k_{t}(\omega)$, which for small frequencies can be written as
\begin{equation}
  k_{t}
	 \approx 
	\sqrt{\omega \epsilon_{0}}\sqrt{\frac{2 \rho \epsilon_{\rm B}}{d_{\rm A}d_{\rm B}}}~,
\label{limitEMA}
\end{equation}
and can be obtained by comparing the results in Eqs. \eqref{eq:taylorslab} and \eqref{rIsuplatt}.
For $k<k_{t}$ the superlattice is well described by the reflection coefficient provided 
by the EMA, while for $k>k_{t}$ the slab description and eventually the bulk description for $k> 1/d_{\rm A}$.

%%%%%%%%

\section{Quantum friction with superlattice structures}
\label{Sec:QFres}

The analyses presented in the previous sections allow for a quantitative assessment of 
quantum friction as well as for a deeper qualitative understanding of the behavior 
of the force in systems involving semi-infinite superlattice substrates. Even if most of the following 
analytical expressions rely on the near-field approximation, the numerical calculations 
consider the entire retarded interaction and are thus exact. 
For the conducting layer, in addition to the 
Drude model in Eq. \eqref{eq:dielfun} with the parameters used to describe gold 
($\omega_\mathrm{p}=9\,\mathrm{eV}$, $\epsilon^\infty=5$ and 
 $\gamma=1.1\times10^{-5} \omega_\mathrm{p}$ \cite{Pirozhenko2008a}), 
we also consider doped silicon 
\begin{eqnarray}
  \epsilon_\mathrm{dop}(\omega)
	 & = &
	\epsilon_\mathrm{Si}(\omega) -\frac{\omega_\mathrm{pSi}^2}{\omega(\omega+i\gamma_\mathrm{Si})}\,.
	\label{eq:dSi}
\end{eqnarray}
In the previous model, the free charge carries are described by 
an additional Drude term, while the intrinsic permittivity of silicon is given by
\begin{eqnarray}
  \epsilon_\mathrm{Si}(\omega)
	  =
	\epsilon_\mathrm{Si}^\infty - \frac{(\epsilon^0_\mathrm{Si}-\epsilon_\mathrm{Si}^\infty)\omega_0^2}{\omega^2-\omega_0^2}
		\label{eq:Si}
\end{eqnarray}
with $\epsilon_\mathrm{Si}^\infty=1.035$, $\epsilon^0_\mathrm{Si}=11.87$ and 
$\omega_0\approx 4.34\,\mathrm{eV}$ \cite{Bergstroem1997}. 
Due to the variability of the doping, we gain access to a wide range of values for the 
resistivity, $\rho_{\rm dSi}=\gamma_\mathrm{Si}/(\epsilon_0\omega_\mathrm{pSi}^2)$,
while maintaining the same basic material description.
The dielectric material B is instead chosen to be intrinsic silicon or, for simplicity, 
 vacuum (i.e. $\epsilon_\mathrm{B}=1$). From the expression presented in the previous 
 section we expect that the value of dielectric function for this layer mostly produces
 a shift or a rescaling of the features induced by the conducting material (see for example Eqs. 
 \eqref{eq:omegapm} and \eqref{rIsuplatt} as well as Fig. \ref{fig:vtrans} and the expression below).

Let us start our analysis by focusing on the non-resonant regime, where 
the force is essentially connected 
to wavevectors $k\lesssim 1/z_{a}$ and frequencies $0<\omega\lesssim v/z_{a}$.
In Sec. \ref{Sec:supEMA}
 we have shown that, depending on the frequencies and the wavevectors, the optical response of superlattice 
structures effectively changes,
featuring behavior typical of a homogeneous bulk, a thin slab or, using the EMA description, an uniaxial crystal. 
Similarly we can expect that, as a function of the atom's velocity and distance from the surface, the 
quantum frictional force explores all the previously discussed regimes. 
%\paragraph{Bulk}

At very low velocities and very short distances, despite the fact that a wide range of frequencies can participate in the interaction,
from the point of view of quantum friction the semi-infinite superlattice behaves as an ohmic medium, indicating a
force which is proportional to $v^{3}$.
The analysis of the previous section, suggests indeed that, as long as $k_{t}z_{a}\ll 1$, the superlattice is equivalent to a metallic bulk or at most a metallic slab. 
In fact, in agreement with the behavior depicted in Fig. \ref{fig:refkp}, for $z_{a}\ll d_{\rm A}$ we recover the expression for the force acting on an atom moving above a homogeneous substrate
composed of an ohmic material \cite{Intravaia2016a}
\begin{eqnarray}
  \bar{F} \sim \bar{F}_\mathrm{bulk}
	  &\stackrel{v\ll c}\approx &
  -\frac{864}{5\pi^3}\hbar\alpha_0^2\rho^2 \frac{v^3}{(2z_a)^{10} }
\,.
\label{eq:asyv}
\end{eqnarray}
As explained above, the $v^3$-scaling is rooted in the linear-in-frequency (ohmic) behavior 
of the imaginary part of reflection coefficient at small frequencies $\omega$ (see Fig. \ref{fig:refbreak}). The $z_a^{-10}$ 
dependence results, instead, from a combined dependence on $k$ and $\omega$ of the total Green 
tensor. For ohmic materials the proportionality to the square of resistivity $\rho$ can be 
directly understood from the functional behavior of Eq.~\eqref{Fsmallv}.
In Eq.~\eqref{eq:asyv} and in all subsequent analytical expressions 
the bar (e.g. $\bar{F}$) indicates the average over all dipole angles.
For simplifying the evaluation, however, our numerical analysis considers the case $\mathbf{d}=\sqrt{\alpha_0/3}\,(1,1,1)$, where 
$\alpha_0=\mathrm{Tr}[\underline{\alpha}_0/3]$.

\begin{figure}
\resizebox{1.05\linewidth}{!}{
\includegraphics[scale=1]{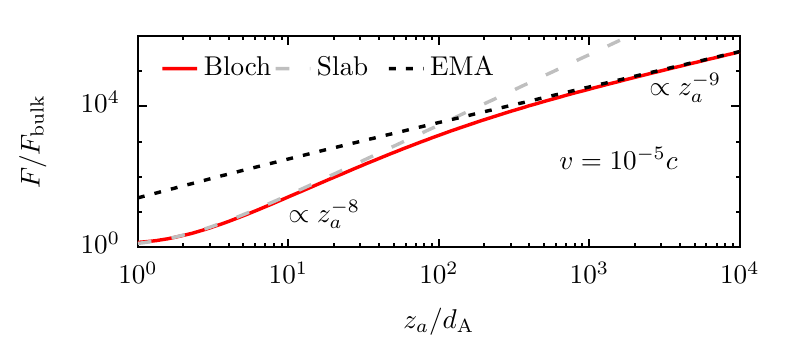}
}
\caption{Quantum frictional force on an atom moving above a superlattice (red solid line) 
         as a function of the atom-surface separation. The force is normalized to the bulk 
				 result [Eq.~\eqref{eq:asyv}]. At large separations, the superlattice can be approximated 
				 by an EMA description (dotted black line) which yields a $z_a^{-9}$ law. At small 
				 distances, instead, the superlattice can essentially be replaced by the top-most 
				 (conducting) layer (gray dashed line). 
         The result for a metallic slab features the transition to bulk behavior, 
				 $z_{a}^{-8}\to z_a^{-10}$ occurring as soon as $z_{a}\ll d_{\rm A}$. As material 
				 parameters we used doped silicon with $\omega_\mathrm{pSi}=0.0725\,\mathrm{eV}$, 
				 $\gamma_\mathrm{Si}=0.0247\,\mathrm{eV}$ and $d_\mathrm{A}=d_\mathrm{B}=1\,\mathrm{nm}$ 
				($\sim 7\times 10^{-12}\,c/\omega_\mathrm{pSi}$). 
\label{fig:ztrans}}
\end{figure}

%\subsection{Slab}

When the distance increases, keeping the low velocity limit, the optical response is still ohmic at low 
frequencies but with a resistivity that effectively increases according to Eq.~\eqref{eq:taylorslab}. 
In this regime, despite the fact that the force still remains proportional to $v^{3}$, the semi-infinite superlattice 
is effectively represented by its first layer and
its thickness, $d_{\rm A}$, appears as an additional length 
scale of the system. This modifies the functional dependence of quantum friction on the atom-surface separation and for $z_{a}\gtrsim d_{\rm A}$
we obtain
\begin{equation}
 \bar{F}\sim\bar{F}_\mathrm{slab}
	\approx
  \bar{F}_\mathrm{bulk} \mathcal{C}_\mathrm{slab}(z_a/d_{\rm A}).
\label{eq:qfzalim}
\end{equation} 
The monotonous and positive function $\mathcal{C}_\mathrm{slab}(x)$, whose explicit form is given in Appendix 
\ref{app1}, is such that $\mathcal{C}_\mathrm{slab}(x\to 0)=1$, recovering the limit of Eq. \eqref{eq:asyv},
and $\mathcal{C}^\mathrm{slab}(x\gg 1)\propto x^2$. Importantly, Eq. \eqref{eq:qfzalim} scales as $z_{a}^{-8}$ for $z_{a}\gtrsim d_{\rm A}$.
Therefore the force decays slower with $z_{a}$ than in Eq. \eqref{eq:asyv}, leading to 
an enhancement of several orders of magnitude with respect to the bulk result
(see Fig.~\ref{fig:ztrans}). In simple terms, this geometry-induced modification and the corresponding increase in the value of the force can be understood as a consequence of the fact that,
while the intrinsic resistivity 
of the material is constant, the layer's resistance effectively increases as its thickness 
is reduced.

%\subsection{EMA}
With a further increase in the atom-surface separation, the changes in the behavior of the quantum frictional force
%Also interesting is the parameter region where 
acting on an atom moving at constant velocity 
above the semi-infinite superlattice become more profound, as
the interaction starts to ``perceive'' the substrate as being well-described by the EMA. According 
to Eq.~\eqref{limitEMA}, in the non-resonant regime
we roughly expect such change of behavior to occur when
\begin{equation}
  vz_{a}\gtrsim \frac{d_{\rm A}d_{\rm B}}{2 \rho \epsilon_{0}\epsilon_{\rm B}},
\label{threshold}
\end{equation}
which also indicates the necessity of sufficiently large velocities (for $v\to0$ the recover the ohmic behavior). In this region we have that
\begin{equation}
  \bar{F}\sim \bar{F}_\mathrm{EMA}\approx - \hbar\alpha_0^2 \frac{6\rho}{\pi^2\epsilon_{0}\epsilon_\mathrm{B}}\frac{d_{\rm B}}{d_{\rm A}}\frac{v|v|}{(2z_a)^{9}}.
\end{equation}
We first notice that the force no longer grows quadratically but linearly with the resistivity 
of the material. The effective sub-ohmic description introduced by the EMA does not only lead 
to a change in the velocity-dependence of the force ($v^{3}\to v^{2}$, as discussed in Sec. \ref{Sec:theoryQF}), 
but also to an additional modification of its functional dependence on the atom-surface separation. 

Figure \ref{fig:ztrans} depicts the quantum frictional force for fixed velocity as a function of the atom-surface separation $z_{a}$. We observe
that with the superlattice structuring, we access the three different regimes discussed above: At short distances, 
we recover the bulk expression $ F\propto z_{a}^{-10}$ given in Eq.~\eqref{eq:asyv}; for
intermediate separations ($z_{a}\gtrsim d_{\rm A}$, the slab regime where $ F\propto 
z_{a}^{-8}$ occurs; finally, for sufficiently large separations, the EMA regime is reached,
yielding $ F\propto z_{a}^{-9}$.

\begin{figure}
\resizebox{1.05\linewidth}{!}{
\includegraphics[scale=1]{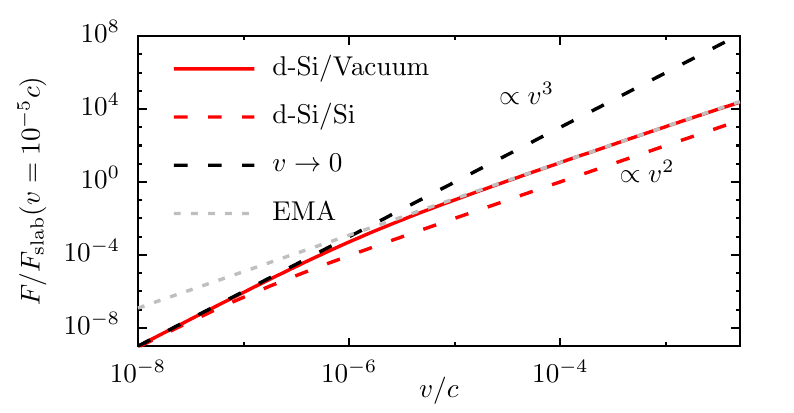}
}
\caption{Velocity dependence of the quantum frictional force, normalized to the value of the 
         slab configuration at $v=10^{-5}\,c$. The atom moves above a superlattice with doped-silicon as conducting layer and intrinsic silicon (solid red line) [see Eq.~\eqref{eq:Si}]  or vacuum (dashed red line) as dielectric layer. The distance is fixed at 
$z_{a}\approx 70\,c/\omega_\mathrm{p}^\mathrm{Si}$ ($\sim 1\mu$m for the materials 
considered). The other parameters are the same of in Fig.~\ref{fig:ztrans}.
In both cases, the full calculations via the Bloch-wave approach
features a transition from $v^{3}$ behavior at low velocities (dashed black line), typical of 
 ohmic materials, to the $v^{2}$ dependence characteristic of the sub-ohmic behavior 
 ($r_{I}\propto  \sqrt{\omega}$) of the EMA (dotted gray line). 
\label{fig:vtrans}}
\end{figure}

\begin{figure}
\resizebox{1.05\linewidth}{!}{
\includegraphics[scale=1]{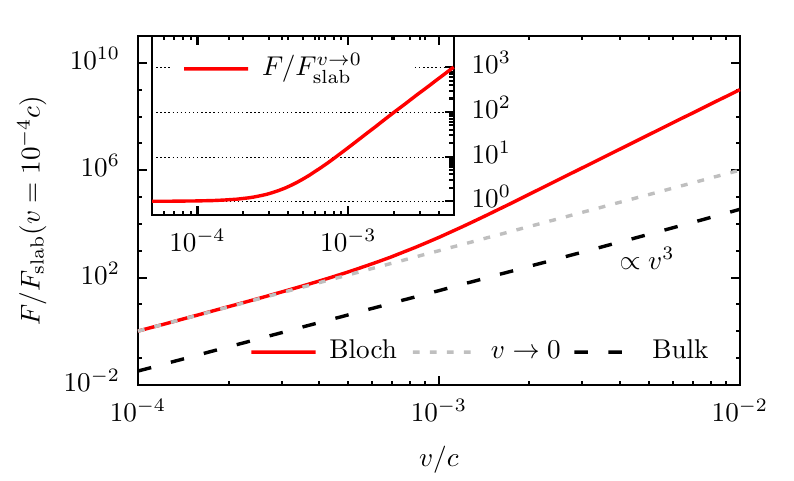}
}
\caption{Resonant enhancement of quantum friction due to the CIPP modes. The quantum frictional 
         force acting on an atom moving above superlattice structures is enhanced due to the 
				 interaction with the coalescence of the CIPP resonances in the $\omega_-$ branch 
				 [Eq. \eqref{eq:omegapm}]. 
				 In order to clearly reveal the effect of CIPP modes, we have chosen a large atomic 
				 transition frequency ($\omega_a=10.2\,\mathrm{eV}$, as e.g. for hydrogen) and a 
				 Drude metal with low damping constant $\gamma=1.1\times10^{-5} \omega_\mathrm{p}$ with 
				 $\omega_\mathrm{p}=9\,\mathrm{eV}$ (further parameters are $\epsilon_\mathrm{A}^\infty=5$, 
				 $d_\mathrm{A}=d_\mathrm{B}=1\,\mathrm{nm}$ and $z_a=10\,\mathrm{nm}$). 
				 For comparison, in addition to the full Bloch-wave calculations (red solid line), 
				 the plot shows the asymptote for small velocities (grey dotted line) and the calculations 
				 for a Drude bulk substrate (dashed black line). The offset for low velocities between 
				 the bulk substrate and the superlattice results can be understood by the different 
				 $z_a$ dependence as displayed in Fig.~\ref{fig:ztrans}. 
				 \textit{Inset}: The full Bloch-wave calculation is normalized by its low-velocity limit, 
				                 which coincides with the result of the slab configuration.
\label{fig:venh}
}
\end{figure}

The velocity dependence of the quantum frictional force is presented in Fig.~\ref{fig:vtrans}. 
At sufficiently low velocities, we recover the $v^3$ law, which is connected to ohmic response 
of the superlattice and this essentially originates from its first layer. However, for increasing 
velocities a broader range of frequencies contributes to the interaction and eventually the 
region where the structure changes its behavior from ohmic to sub-ohmic becomes relevant. This 
corresponds to a change of the velocity dependence of the force from $v^3$ 
to $v^2$.
Finally, it is interesting to consider the characteristic of the resonant contribution to quantum friction in systems
involving semi-infinite superlattices.
As described in Sec.~\ref{Sec:supEMA} and shown 
in Fig. \ref{fig:refbreak}, for certain parameters we observe large values of $r_{I}$ 
due to the coalescence of the resonances occurring in the $\omega_{\pm}$ branches. This behavior, which
is connected with the hyperbolic dispersion of the structure, is particularly 
evident for a superlattice composed of low-damping materials, where the sub-ohmic regime is 
less pronounced. For these frequencies, even if the material has very weak dissipation,
the continuum of modes in the $\omega_-$ branch (and similarly but at higher frequencies for the $\omega_+$ branch) effectively behaves as an ``energy sink'' which, through a non-radiative coupling with the atom, can efficiently transport energy away from the surface through the superlattice.
Depending on the velocity and the distance of the atom, this frequency region 
can give rise to a resonant contribution, which leads to an additional increase of the force. 
The interaction generating the CIPP also shifts this frequency range to a frequency below 
$\omega_{\rm sp}$, lowering the corresponding resonant velocity threshold and adding a certain
degree of tunability via the thickness of the layers. In Fig.~\ref{fig:venh}, we indeed observe 
a steady increase of the quantum frictional force that occurs at a relatively low velocity.
Due to the broadband nature of the $\omega_{-}$ band, this resonant contribution differs from 
that generated by an isolated resonance (see for example Refs. \cite{Intravaia2016,Intravaia2016a}) 
and the system features a smoother transition out of the non-resonant regime. 

\section{Conclusions}

Modern technologies allow for the structuring of materials at the size of nanometers, 
prompting novel applications in several areas of physics. In this work, we have shown 
that such nanostructuring can be very interesting with regards to nonequilibrium 
atom-surface interactions and in particular for controlling the strength and the 
functional dependencies of the quantum frictional force on an atom moving at constant 
velocity and height above multi-layered structures. Indeed, when these structures
consist of a semi-infinite superlattice of alternating metallic and dielectric layers, 
the spectrum of vacuum fluctuations is considerably modified relative to that of an 
homogeneous medium and can be tuned by changing the thickness and the material properties 
of the layers. 
In these systems, the frequency spectrum is characterized by the appearance of coupled 
interface plasmon-polariton (CIPP) modes: They arise from the electromagnetic interaction 
among the charge-carrier densities existing at the metal-dielectric interfaces and can 
be considered as the generalization of the surface plasmon-polariton resonances appearing 
at metallic surfaces. Mathematically, for semi-infinite superlattices, the CIPP modes 
manifest themselves as two continua characterized by well-prescribed symmetries of the 
associated electromagnetic field.
Their behavior is also connected with the properties
of the superlattice to exhibit hyperbolic dispersions.
We have seen that CIPP modes affect the quantum frictional forces in different ways 
depending on the speed of the atom and on its distance from the surface. At low velocity, 
the force is strongly connected with the low-frequency behavior of the surface's 
$p$-polarized reflection coefficient, $r^{p}(\omega, k)$. In superlattices, depending 
on the wavevector, we have observed a change in behavior of the imaginary part of the 
reflection coefficient, from ohmic ($r^{p}_{I}\propto \omega$) to sub-ohmic ($r^{p}_{I}\propto \sqrt{\omega}$). 
The former can be ascribed to the electromagnetic response of the first metallic layer 
in the stacking sequence, while the latter is related to an overdamped (non-resonant) subset of the
CIPP modes. We have shown that, for the quantum frictional force, this behavior can 
lead to enhancements of several orders of magnitude relative to the case of a homogeneous 
bulk substrate, as well as to a modification of the power law describing its functional 
dependence on the atom-surface separation ($z_a^{-10}\to z_a^{-8}\to z_{a}^{-9}$, see 
Fig.~\ref{fig:ztrans}). 
Similarly, the velocity dependence changes from $\propto v^{3}$, typical of the ohmic 
response, to $\propto v^{2}$ induced by $r^{p}_{I}\propto \sqrt{\omega}$ (see 
Fig.~\ref{fig:vtrans}). The threshold distances and velocities, where these 
transitions take place, depend on the material properties and geometrical parameters 
of the system [Eq.~\eqref{threshold}].
Finally, at higher velocities resonant phenomena can become important. An analysis of 
the relevant expressions shows that the velocity threshold, when this occurs, is usually
rather high due to the large typical values of the involved resonance 
frequencies. However, we have shown that in superlattice systems, by reducing the layers' 
thickness, the interaction among all the surface plasmon polaritons at the different
material interfaces lead to a displacement of the continuum of CIPP resonances to 
lower energy, allowing for a more accessible resonant enhancement of the quantum 
frictional force (see Fig.~\ref{fig:venh}). This behavior has been further connected with the
hyperbolic properties of the semi-infinite superlattice. Indeed even for low dissipative materials,
the increase in the density of states connected with the hyperbolic dispersion creates an additional
channel through which energy can be carried deep into the substrate.

These results highlight once more the role of geometry and material properties in 
fluctuation-induced phenomena and indicate a pathway for future experimental investigations 
of nonequilibrium atom-surface effects. While the geometry can be used to control 
(enhance or suppress) the interaction by changing the functional dependence of the
quantum frictional force, the material properties offer a direct access to the 
proportionality constants. As an example, using high-resistivity materials such as
GaAs ($\rho_\mathrm{GaAs}\approx 10^{9}\,\Omega\mathrm{cm}$) in a superlattice 
structure (vacuum as a dielectric, for simplicity) with $10\,\mathrm{nm}$ thick 
layers, our analysis predicts a quantum friction force of $F\approx-15\,\mathrm{fN}$ 
acting on a $^{87}\mathrm{Rb}$ atom moving at a height of $z_a=0.1\,\mathrm{\mu m}$ 
above the multi-layer surface with a velocity of $v=5\times 10^{-4} c$. This value 
of the quantum frictional force corresponds to an acceleration of about $10^{11}\mathrm{m/s^{2}}$
which is within the presently available experimentally measurable accuracy. 
An experimental confirmation of quantum frictional forces would be of high fundamental 
interest and can provide a deeper understanding of the underlying physics of 
nonequilibrium quantum-fluctuation-induced phenomena.

\section*{Acknowledgement}
We would like to thank D. Reiche, D. Huynh and Ch. Egerland for useful and fruitful 
discussions. IN addition, we acknowledge support by the Deutsche Forschungsgemeinschaft 
(DFG) through project B10 within the Collaborative Research Center (CRC) 951 Hybrid 
Inorganic/Organic opto-electronic Systems(HIOS). FI further acknowledges financial 
support from the DFG through the DIP program (Grant No. SCHM 1049/7-1).

\appendix
\section{Definitions and low-velocity limit}
\label{app1}
The definition of the quantum frictional force essentially depends on two susceptibilities,
the atomic polarizability, characterizing the moving microscopic object, and the Green 
tensor that characterizes the electromagnetic properties of the nanostructured substrate. 
In general, the Green tensor can be written as the sum of the vacuum contribution 
$\underline{G}_0$ and a scattered contribution $\underline{g}$. While the expression 
for the former can be found in textbooks (see for example Ref.~\cite{Jackson75}), for
flat surfaces $\underline{g}$ the latter takes the form \cite{Wylie1984}
\begin{equation}
  \underline{g}(\mathbf{k},z_a,\omega)
	 =
  \frac{e^{-2\kappa z_a}\kappa}{2\epsilon_0}\left[ r^p(\omega,k) \mathbf{p}_+\mathbf{p}_-
                                                   +
                                                   \frac{\omega^2}{c^2\kappa^2}r^s(\omega,k) \mathbf{s}\mathbf{s}
                                            \right]~,
\label{eq:greene}
\end{equation}
Here, we have introduced $\kappa=\sqrt{k^2-\omega^2/c^2}$ ($\mathrm{Re\lbrace\kappa\rbrace>0}$ 
and $\mathrm{Im\lbrace\kappa\rbrace<0}$). Further, $k^2=k_x^2+k_y^2$ and $r^\sigma$ are the 
reflection coefficients for the two polarizations, $\sigma=s,p$, and 
\begin{equation} 
  \mathbf{s}
	  =
	\frac{\mathbf{k}}{k}\times\frac{\mathbf{z}}{z}, \quad
  \mathbf{p}_\pm
	  =
	\frac{k}{\kappa}\frac{\mathbf{z}}{z}\mp i\frac{\mathbf{k}}{k}\,.
\end{equation}
In these expressions, $\mathbf{z}$ represents the vector in $z$ direction (perpendicular 
to the surface of the substrate).
In terms of the Green tensor we can also define the velocity-dependent polarizability tensor
\begin{eqnarray}
  \underline{\alpha}(\omega,\mathbf{v})
	 & = &
  \frac{\underline{\alpha}_0\omega_a^2}{\omega_a^2-\omega^2-\Delta_a(\omega;\mathbf{v})-i\omega\gamma_a(\omega;\mathbf{v})},
\end{eqnarray}
where the dyadic $\underline{\alpha}_0=\mathbf{d}\mathbf{d}$ is the static polarizability 
and have introduced the abbreviations
\begin{subequations}
\begin{equation}
  \Delta_a(\omega) 
	  = 
	\omega_a^2\int\frac{\mathrm{d}^2\mathbf{k}}{(2\pi)^2}\,
	\mathrm{Tr}\left[ \underline{\alpha}_0
                    \cdot
                    \underline{G}_R(\mathbf{k},z_a,\omega+\mathbf{k}\cdot\mathbf{v})
             \right],
\end{equation}
and
\begin{equation}
  \gamma_a(\omega) 
	  = 
	\frac{\omega_a^2}{\omega}\int\frac{\mathrm{d}^2\mathbf{k}}{(2\pi)^2}\,
	\mathrm{Tr}\left[ \underline{\alpha}_0                 
                    \cdot
                    \underline{G}_I(\mathbf{k},z_a,\omega+\mathbf{k}\cdot\mathbf{v})
             \right]
\end{equation}
\end{subequations}
which, respectively, describe the induced frequency shift and damping. The polarizability also 
appears in the expression of the nonequilibrium correction to the fluctuation-dissipation 
theorem
\begin{eqnarray}
  \underline{J}(\omega; \mathbf{v}) 
	 & = &
  \int \frac{\mathrm{d}^2\mathbf{k}}{(2\pi)^2}
	\left[ \theta(\omega+\mathbf{k}\cdot\mathbf{v}) - \theta( \omega) \right] \\
	 &   &
	\times\, \underline{\alpha}(\omega; \mathbf{v})\cdot \underline{G}_I(\mathbf{k},z_a,\omega+\mathbf{k}\cdot\mathbf{v}) \cdot\underline{\alpha}^*(\omega; \mathbf{v})\nonumber
\label{eq:J}
\end{eqnarray}
which occurs in Eq.~\eqref{PS}.

Altogether, the above expressions allow the evaluation of the non-relativistic value of 
the quantum frictional force given in Eq.~\eqref{eq:QFint}. The structure of Eq.~\eqref{PS} 
indicates that the quantum friction force can be decomposed into a contribution related to 
the local thermal equilibrium (LTE) approximation and a full nonequilibrium correction. 
This separation is also visible in low velocity approximation of the force given in 
Eq.~\eqref{Fsmallv}, where the first term on the r.h.s. is the result within the LTE, 
$F^\mathrm{LTE}$, while the second, $F^J$, is entirely due to the the tensor 
$\underline{J}(\omega; \mathbf{v})$. 

Equation \eqref{Fsmallv} also shows that the low-velocity behavior of the force is connected 
to the low-frequency features of the nano-structures optical response and eventually with 
the low-frequency expansion of the imaginary part of the reflection coefficients. We have
seen in the main text that for metal-dielectric superlattice structures, at sufficiently 
small atom-surface separations, the optical response is dominated by the first (metallic) 
layer. Effectively, the quantum frictional force felt by the atom is the same as that 
produced by a metallic slab, i.e., $F_\mathrm{sup}\approx F_\mathrm{slab}$. Using the 
expressions in Eq.~\eqref{eq:taylorslab} this allows for the following estimates. For the 
LTE term, we obtain
\begin{multline}
  \frac{F^\mathrm{LTE}_\mathrm{slab}}{F_\mathrm{bulk}^\mathrm{LTE}}
	  =
	\mathcal{C}^\mathrm{LTE}_\mathrm{slab}(\tfrac{z_a}{D})\\
    \stackrel{v\rightarrow0}\approx
  \frac{\int_0^\infty \mathrm{d}k\,k^6 e^{-2kz_a}\coth(kD)}{\int_0^\infty \mathrm{d}k\,k^6 e^{-2kz_a}} \\
  \times\frac{\int_0^\infty \mathrm{d}k\, k^2 e^{-2kz_a}\coth(kD)}{\int_0^\infty \mathrm{d}k\, k^6 e^{-2kz_a}}
\end{multline}
where $F_\mathrm{bulk}^\mathrm{LTE}$ is the LTE contribution to the quantum frictional force 
in the case of a homogeneous semi-infinite substrate. Its value 
\begin{equation}
  \bar{F}_\mathrm{bulk}^\mathrm{LTE}
    \approx
   -\frac{21}{20}\frac{90}{\pi^3}\hbar\alpha_0^2\rho^2\frac{v^3}{(2z_a)^{10}}
\end{equation}
has already ben calculated in Ref.~\cite{Intravaia2016a}, where the bar indicates the average 
over the dipole angles. The function that gives the correction induced by the finite thickness 
is defined as 
\begin{equation}
  \mathcal{C}_\mathrm{slab}^\mathrm{LTE}(\tfrac{z_a}{D})
	  =
	\left[ 1 - 2\frac{\zeta\left(7,\frac{z_a}{D}\right)}{(D/z_a)^7} \right]
	\left[ 1 - 2\frac{\zeta\left(3,\frac{z_a}{D}\right)}{(D/z_a)^3} \right]\,,
\end{equation}
where 
\begin{equation}
  \zeta(s,x)
	  =
	\sum_{n=0}^{\infty}\frac{1}{(n+x)^{s}} 
\end{equation}
is the Hurwitz Zeta function. Upon applying the same strategy to the nonequilibrium contribution 
we obtain an analogous expression that reads as
\begin{equation}
  \bar{F}_\mathrm{bulk}^{J}
	  \approx 
  -\frac{87}{80}\frac{72}{\pi^3}\hbar\alpha_0^2\rho^2\frac{v^3}{(2z_a)^{10}}
\end{equation}
and the corresponding correction function
\begin{equation}
  \mathcal{C}_\mathrm{slab}^{J}(\tfrac{z_a}{D})
	  =
	\left[1 -2\frac{\zeta\left(5,\frac{z_a}{D}\right)}{(D/z_a)^5} \right]^2\,.
\end{equation}
Adding the two contributions, we can define the total correction function introduced in 
Eq. \eqref{eq:qfzalim} 
\begin{equation}
  \mathcal{C}_\mathrm{slab}(\tfrac{z_a}{D})
	  =
	\frac{F^\mathrm{LTE}_\mathrm{bulk}\mathcal{C}_\mathrm{slab}^\mathrm{LTE} (\tfrac{z_a}{D})
	+ 
	F^J_\mathrm{bulk}\mathcal{C}_\mathrm{slab}^{J}(\tfrac{z_a}{D})}{F^\mathrm{LTE}_\mathrm{bulk}+ F^J_\mathrm{bulk}},
\end{equation}
which clearly inherits the properties of the functions defined above.
For $z_a \gg D$ we have that
\begin{align}
  \bar{F}
	 & \approx 
	\bar{F}_\mathrm{slab}\approx \stackrel{z_a \gg D}
	   \approx
	-\frac{\hbar \alpha_0^2\rho^2}{D^2\pi^3}\frac{v^3}{(2z_a)^{8}}\frac{2043}{160}.
\label{eq:supfric}
\end{align}

%\bibliography{literature,/Users/nabu/Documents/Mydocs/Lavoro/bibliography/biblio}

\begin{thebibliography}{10}

\bibitem{Casimir1948}
H.~B.~G. Casimir and D. Polder, The Influence of Retardation on the London-van
  der Waals Forces, Phys. Rev. {\bf 73},  360  (1948).

\bibitem{Intravaia2016}
F. Intravaia, R.~O. Behunin, C. Henkel, K. Busch, and D.~A.~R. Dalvit,
  Non-Markovianity in atom-surface dispersion forces, Phys. Rev. A {\bf 94},
  042114  (2016).

\bibitem{Intravaia15}
F. Intravaia, V.~E. Mkrtchian, S.~Y. Buhmann, S. Scheel, D.~A.~R. Dalvit, and
  C. Henkel, Friction forces on atoms after acceleration, J. Phys. Condens.
  Matter {\bf 27},  214020  (2015).

\bibitem{Pieplow15}
G. Pieplow and C. Henkel, Cherenkov friction on a neutral particle moving
  parallel to a dielectric, J. Phys. Condens. Matter {\bf 27},  214001  (2015).

\bibitem{Dedkov02a}
G. Dedkov and A. Kyasov, Electromagnetic and fluctuation-electromagnetic forces
  of interaction of moving particles and nanoprobes with surfaces: A
  nonrelativistic consideration, Phys. Solid State {\bf 44},  1809  (2002).

\bibitem{Scheel2009}
S. Scheel and S. Y. Buhmann, Casimir-Polder forces on moving atoms,
Phys. Rev. A {\bf 80}, 042902 (2009).

\bibitem{Yan2012}
W. Yan, M. Wubs, and N.~A. Mortensen, Hyperbolic metamaterials: Nonlocal
  response regularizes broadband supersingularity, Phys. Rev. B {\bf 86},
  205429  (2012).

\bibitem{Liu2007}
Z. Liu, H. Lee, Y. Xiong, C. Sun, and X. Zhang, Far-Field Optical Hyperlens
  Magnifying Sub-Diffraction-Limited Objects, Science {\bf 315},  1686  (2007).

\bibitem{Shekhar2014}
P. Shekhar, J. Atkinson, and Z. Jacob, Hyperbolic metamaterials: fundamentals
  and applications, Nano Convergence {\bf 1},  1  (2014).

%\bibitem{Davids10}
%P.~S. Davids, F. Intravaia, F.~S.~S. Rosa, and D.~A.~R. Dalvit, Modal approach
%  to Casimir forces in periodic structures, Phys. Rev. A {\bf 82},  062111
%  (2010).

\bibitem{Rodriguez11}
A.~W. Rodriguez, F. Capasso, and S.~G. Johnson, The Casimir effect in
  microstructured geometries, Nat. Photon. {\bf 5},  211  (2011).

\bibitem{Intravaia12a}
F. Intravaia, P.~S. Davids, R.~S. Decca, V.~A. Aksyuk, D. L\'opez, and D.~A.~R.
  Dalvit, Quasianalytical modal approach for computing Casimir interactions in
  periodic nanostructures, Phys. Rev. A {\bf 86},  042101  (2012).


\bibitem{Lambrecht2008}
A. Lambrecht and V. N. Marachevsky, Casimir Interaction of Dielectric Gratings,
Phys. Rev. Lett. {\bf 101}, 160403 (2008).


\bibitem{Messina2017}
R. Messina, A. Noto, B. Guizal, and M. Antezza,
Radiative heat transfer between metallic gratings using Fourier modal method with adaptive spatial resolution,
Phys. Rev. B \textbf{95}, 125404 (2017)


\bibitem{Chan08}
H.~B. Chan, Y. Bao, J. Zou, R.~A. Cirelli, F. Klemens, W.~M. Mansfield, and
  C.~S. Pai, Measurement of the Casimir Force between a Gold Sphere and a
  Silicon Surface with Nanoscale Trench Arrays, Phys. Rev. Lett. {\bf 101},
  030401  (2008).

\bibitem{Intravaia13}
F. Intravaia {\it et~al.}, Strong Casimir force reduction through metallic
  surface nanostructuring, Nat. Commun. {\bf 4},  2515  (2013).

\bibitem{Bender14}
H. Bender, C. Stehle, C. Zimmermann, S. Slama, J. Fiedler, S. Scheel, S.~Y.
  Buhmann, and V.~N. Marachevsky, Probing Atom-Surface Interactions by
  Diffraction of Bose-Einstein Condensates, Phys. Rev. X {\bf 4},  011029
  (2014).

\bibitem{Tang17}
L. Tang, M. Wang, C.~Y. Ng, M. Nikolic, C.~T. Chan, A.~W. Rodriguez, and H.~B.
  Chan, Measurement of non-monotonic Casimir forces between silicon
  nanostructures, Nat Photon {\bf 11},  97  (2017).

\bibitem{Chan18}
E.~A. Chan, S.~A. Aljunid, G. Adamo, A. Laliotis, M. Ducloy, and D. Wilkowski,
  Tailoring optical metamaterials to tune the atom-surface Casimir-Polder
  interaction, Sci. Adv. {\bf 4},    (2018).

\bibitem{Saarinen2008}
J.~J. Saarinen, S.~M. Weiss, P.~M. Fauchet, and J.~E. Sipe, Reflectance
  analysis of a multilayer one-dimensional porous silicon structure: Theory and
  experiment, Journal of Applied Physics {\bf 104},  013103  (2008).

\bibitem{Poddubny13}
A. Poddubny, I. Iorsh, P. Belov, and Y. Kivshar, Hyperbolic metamaterials, Nat.
  Photon. {\bf 7},  948  (2013).

\bibitem{Guo14b}
Y. Guo and Z. Jacob, Fluctuational electrodynamics of hyperbolic metamaterials,
  J. Appl. Phys. {\bf 115},    (2014).

\bibitem{Iorsh12}
I. Iorsh, A. Poddubny, A. Orlov, P. Belov, and Y.~S. Kivshar, Spontaneous
  emission enhancement in metal--dielectric metamaterials, Phys. Lett. A {\bf
  376},  185   (2012).

\bibitem{Kidwai12}
O. Kidwai, S.~V. Zhukovsky, and J.~E. Sipe, Effective-medium approach to
  planar multilayer hyperbolic metamaterials: Strengths and limitations, Phys.
  Rev. A {\bf 85},  053842  (2012).


\bibitem{Belov06}
P.~A. Belov and Y. Hao, Subwavelength imaging at optical frequencies using
  a transmission device formed by a periodic layered metal-dielectric structure
  operating in the canalization regime, Phys. Rev. B {\bf 73},  113110
  (2006).

\bibitem{Wurtz11}
G.~A. Wurtz, R. Pollard, W. Hendren, G.~P. Wiederrecht, D.~J. Gosztola, V.~A.
  Podolskiy, and A.~V. Zayats, Designed ultrafast optical nonlinearity in
  a plasmonic nanorod metamaterial enhanced by nonlocality, Nat Nano {\bf 6},
  107  (2011).

\bibitem{Cortes12}
C.~L. Cortes, W. Newman, S. Molesky, and Z. Jacob, Quantum nanophotonics
  using hyperbolic metamaterials, J. Optics {\bf 14},  063001  (2012).

\bibitem{Intravaia2014}
F. Intravaia, R.~O. Behunin, and D.~A.~R. Dalvit, Quantum friction and
  fluctuation theorems, Phys. Rev. A {\bf 89},  050101  (2014).

\bibitem{Intravaia2016a}
F. Intravaia, R.~O. Behunin, C. Henkel, K. Busch, and D.~A.~R. Dalvit, Failure
  of Local Thermal Equilibrium in Quantum Friction, Phys. Rev. Lett. {\bf 117},
   100402  (2016).

\bibitem{Callen1951}
H.~B. Callen and T.~A. Welton, Irreversibility and Generalized Noise, Phys.
  Rev. {\bf 83},  34  (1951).

\bibitem{Intravaia14}
F. Intravaia, R.~O. Behunin, and D.~A.~R. Dalvit, Quantum friction and
  fluctuation theorems, Phys. Rev. A {\bf 89},  050101(R)  (2014).

\bibitem{Reiche2017}
D. Reiche, D.~A.~R. Dalvit, K. Busch, and F. Intravaia, Spatial dispersion in
  atom-surface quantum friction, Phys. Rev. B {\bf 95},  155448  (2017).

\bibitem{Maghrebi13a}
M.~F. Maghrebi, R. Golestanian, and M. Kardar, Quantum Cherenkov radiation and
  noncontact friction, Phys. Rev. A {\bf 88},  042509  (2013).


\bibitem{Ford84}
G.~W. Ford and W.~H. Weber, Electromagnetic interactions of molecules with
  metal surfaces, Phys. Rep. {\bf 113},  195  (1984).

\bibitem{Note1}
In order to focus on the nano-structuring, we neglect the influence of \protect
  \textit {non-local} effects in material properties throughout all the paper.
  For a detailed study of non-locality in this context of quantum friction see
  Ref.\cite {Reiche2017}.


\bibitem{Sipe1981}
J. Sipe, The dipole antenna problem in surface physics: A new approach, Surface
  Science {\bf 105},  489   (1981).

\bibitem{Jackson75}
J. Jackson, {\em Classical Electrodynamics} (John Wiley and Sons Inc., New
  York, 1975).

\bibitem{Camley1984}
R.~E. Camley and D.~L. Mills, Collective excitations of semi-infinite
  superlattice structures: Surface plasmons, bulk plasmons, and the
  electron-energy-loss spectrum, Phys. Rev. B {\bf 29},  1695  (1984).

\bibitem{Berini00}
P. Berini, Plasmon-polariton waves guided by thin lossy metal films of finite
  width: Bound modes of symmetric structures, Phys. Rev. B {\bf 61},  10484
  (2000).

\bibitem{Burke1986}
J.~J. Burke, G.~I. Stegeman, and T. Tamir, Surface-polariton-like waves guided
  by thin, lossy metal films, Phys. Rev. B {\bf 33},  5186  (1986).

\bibitem{Berini09}
P. Berini, Long-range surface plasmon polaritons, Adv. Opt. Photon. {\bf 1},
  484  (2009).

\bibitem{Barchiesi2014}
D. Barchiesi and T. Grosges, Fitting the optical constants of gold, silver,
  chromium, titanium, and aluminum in the visible bandwidth, Journal of
  Nanophotonics {\bf 8},  083097  (2014).

\bibitem{Albuquerque2004}
E.~L. Albuquerque and M.~G. Cottam, {\em Polaritons in Periodic and
  Quasiperiodic Structures} (Elsevier Science, Amsterdam, 2004).

\bibitem{Yariv1984}
A. Yariv and P. Yeh, {\em Optical Waves in Crystals: Propagation and Control of
  Laser Radiation}, {\em Wiley Series in Pure and Applied Optics} (Wiley,
  New York, 1984).

\bibitem{Intravaia2015}
F. Intravaia and K. Busch, Fluorescence in nonlocal dissipative periodic
  structures, Phys. Rev. A {\bf 91},  053836  (2015).

\bibitem{Bloch1929}
F. {Bloch}, {{\"U}ber die Quantenmechanik der Elektronen in Kristallgittern},
  Zeitschrift f\"ur Physik {\bf 52},  555  (1929).


\bibitem{Mochan1987}
W.~L. Moch\'an, M. del Castillo-Mussot, and R.~G. Barrera, Effect of plasma
  waves on the optical properties of metal-insulator superlattices, Phys. Rev.
  B {\bf 35},  1088  (1987).

\bibitem{Barton79}
G. Barton, Some surface effects in the hydrodynamic model of metals, Rep. Prog.
  Phys. {\bf 42},  963  (1979).

\bibitem{Note2}
In addition to the CIPP, when the thickness of the metallic layer is larger
  than that of the dielectric layer ($d_{\protect \rm A}>d_{\protect \rm B}$),
  additional surface modes can appear in the electromagnetic spectrum characterizing
  the system \cite {Camley1984}. For simplicity we will exclude them from the
  present investigation, by limiting our analysis to the case $d_{\protect \rm
  A}\le d_{\protect \rm B}$.

\bibitem{Note3}
In most of the calculations we are interested in the low-frequency behavior of the functions involved in the evaluation of quantum friction. In this limit the optical response of a dielectric is described with good approximation by a real positive constant constant, i.e. $\epsilon_{\rm B}(\omega)\sim \epsilon_{\rm B}>0$.


\bibitem{chebykin11}
A.~V. Chebykin, A.~A. Orlov, A.~V. Vozianova, S.~I. Maslovski, Y.~S. Kivshar,
  and P.~A. Belov, Nonlocal effective medium model for multilayered
  metal-dielectric metamaterials, Phys. Rev. B {\bf 84},  115438  (2011).

\bibitem{Johnson1985}
B.~L. Johnson, J.~T. Weiler, and R.~E. Camley, Bulk and surface plasmons and
  localization effects in finite superlattices, Phys. Rev. B {\bf 32},  6544
  (1985).

\bibitem{Bruggeman1935}
D.~A.~G. Bruggeman, Berechnung verschiedener physikalischer Konstanten von
  heterogenen Substanzen. I. Dielektrizit\"{a}tskonstanten und Leitf\"{a}higkeiten
  der Mischk\"{o}rper aus isotropen Substanzen, Annalen der Physik {\bf 416},  636
   (1935).

\bibitem{Landau84}
L. Landau and E. Lifshitz, {\em Electrodynamics of Continuous Media}, Vol.~8 of
  {\em Course of Theoretical Physics}, second edition revised and enlarged ed.
  (Pergamon, Amsterdam, 1984).

\bibitem{Knoesen1985}
A. Knoesen, M.~G. Moharam, and T.~K. Gaylord, Electromagnetic propagation at
  interfaces and in waveguides in uniaxial crystals, Applied Physics B {\bf
  38},  171  (1985).

\bibitem{Isic2017}
G. Isi\'{c}, S. Vukovi\'{c}, Z. Jak\v{s}i\'{c} and M. Beli\'{c}, 
Tamm plasmon modes on semi-infinite metallodielectric superlattices,
Scientific Reports \textbf{7}, 3746 (2017).

\bibitem{Pirozhenko2008a}
I. Pirozhenko and A. Lambrecht, Influence of slab thickness on the Casimir
  force, Phys. Rev. A {\bf 77},  013811  (2008).

\bibitem{Bergstroem1997}
L. Bergstr\"om, Hamaker constants of inorganic materials, Advances in Colloid
  and Interface Science {\bf 70},  125   (1997).

\bibitem{Wylie1984}
J.~M. Wylie and J.~E. Sipe, Quantum electrodynamics near an interface, Phys.
  Rev. A {\bf 30},  1185  (1984).


\end{thebibliography}
%\bibliographystyle{/Users/nabu/Documents/Mydocs/Lavoro/bibliography/bibstyle/prstytitlenew}

\end{document}